\documentclass[aps,twocolumn,reprint,preprintnumbers,showpacs]{revtex4-2}
\usepackage[english]{babel}

\usepackage{booktabs}
\usepackage{amsfonts}

\usepackage{tikz,subfigure}
\usepackage{amsmath}
\usepackage{graphicx}
\usepackage[colorlinks=true, allcolors=blue]{hyperref}
\begin{document}
\preprint{MPP-2025-228, USTC-ICTS/PCFT-25-62, LAPTH-055/25}
\title{Three-loop pentagonal Wilson loop with Lagrangian insertion}
\author{Dmitry Chicherin$^{a}$}
\email{chicherin@lapth.cnrs.fr}
\author{Johannes Henn$^{b}$}
\email{Corresponding author. E-Mail: henn@mpp.mpg.de.}
\author{Yongqun Xu$^{c}$}
\email{yongqunxu@mail.ustc.edu.cn}
\author{Shun-Qing Zhang$^{b}$}
\email{sqzhang@mpp.mpg.de}
\author{Yang Zhang$^{c,d,e}$}
\email{yzhphy@ustc.edu.cn}

\affiliation{
$a$ LAPTh, Universit\'e Savoie Mont Blanc, CNRS, B.P. 110, F-74941 Annecy-le-Vieux, France\\
$^b$ Max-Planck-Institut f\"{u}r Physik, Werner-Heisenberg-Institut, Boltzmannstr. 8,
85748 Garching, Germany \\
$^c$ Interdisciplinary Center for Theoretical Study, University of Science and Technology of China, Hefei, Anhui 230026, China\\
$d$ Peng Huanwu Center for Fundamental Theory, Hefei, Anhui 230026, China\\
$e$ Center for High Energy Physics, Peking University, Beijing 100871, People’s Republic of China
}
\begin{abstract}
Employing a cutting-edge bootstrap method, we analytically compute the three-loop pentagonal Wilson loop with Lagrangian insertion in planar $\mathcal{N}=4$ super-Yang-Mills theory. This object is conjectured to coincide with the maximally transcendental part of the four-loop five-point all-plus amplitude in pure Yang-Mills theory. 
Our starting point is an ansatz that encodes the known
leading singularities of this object, as well as the relevant function space. The latter has become available only recently, thanks to an analytic computation of all three-loop five-point planar massless Feynman integrals. 
 We determine the coefficients in the ansatz by imposing physical constraints. This includes a near-collinear expansion, which so far has not been applied to this observable. 
 Taken together, the constraints allow us to uniquely determine the symbol of the answer.
 We verify the symbol result by an independent integral reduction calculation.
\end{abstract}

\maketitle

\section{Introduction}

Wilson loops play a crucial role in the study of gauge theories. 
All gauge-invariant information can be expressed  in terms of Wilson loops. Moreover, Wilson loops are ubiquitous in QCD as an effective physical description. Examples are soft-gluon resummation, jet physics, and factorization theorems.
In planar maximally super-Yang-Mills theory (sYM),
a surprising duality between scattering amplitudes and polygonal Wilson loops has been central to many studies of scattering amplitudes in this theory, such as the discovery of hidden symmetries, a novel operator product expansion, analytic insights into the relevant function space combined with symbol bootstrap approaches, novel geometric amplitude representations.
For recent reviews, see refs. \cite{Henn:2020omi,Arkani-Hamed:2022rwr}.
More recently, it was discovered that a generalization of those Wilson loops---adding a Lagrangian insertion---conjecturally makes a connection to scattering amplitudes in pure Yang-Mills theory \cite{Chicherin:2022bov}. This opens up the exciting possibility of taking inspiration from the powerful developments in sYM, and learning from them for non-supersymmetric scattering amplitudes.

We study the following ratio of vacuum expectation values in sYM theory,
\begin{align}\label{eq:defFintro}
F_n(x_1, \ldots, x_n; x_0) 
= \pi^2 \frac{\langle  W_n \mathcal{L}(x_0)  \rangle}{\langle  W_n  \rangle}\,,
\end{align}
where the Wilson loops $W_n$ are defined for a null polygon with cusp points $x_1, \ldots, x_n$. The edges of this polygon are the momenta of the dual scattering amplitudes.
In sYM, $F_n$ is both infrared and ultraviolet finite, and therefore can be computed directly in four dimensions.
Its integrand is known in principle to all loops from the Amplituhedron \cite{Arkani-Hamed:2013jha}. Moreover, the full set of leading singularities was initially conjectured in refs. \cite{Chicherin:2022bov,Chicherin:2022zxo}, and proven in ref. \cite{Brown:2025plq}.
The maximally transcendental part of $F_n$ is conjectured \cite{Chicherin:2022bov} to be dual to the all-plus-helicity amplitude in pure Yang-Mills theory.
Thanks to this connection, obtaining  perturbative results for $F_n$ also teaches us about the function space needed for Yang-Mills scattering amplitudes.

An attractive way to learn about physical quantities is via the {\it{symbol bootstrap}}. In this method, one sets up an ansatz for the answer. The main ingredients are the expected leading singularities and symbols of the transcendental functions. 
In ref. \cite{Carrolo:2025pue}, the two-loop six-point contribution was bootstrapped in this way, leveraging new insights into the function space \cite{Abreu:2024fei, Henn:2025xrc}. 
 However, despite progress \cite{Chicherin:2024hes}, determining the three-loop five-point case remained an important challenge.
This is due to the previously unknown function space, and due to the technical complexity of handling a large number of multi-variable functions.
Recent results on the three-loop five-point function space, cf. ref. \cite{Chicherin:2025}, have now opened up a new opportunity. 
In this {\it Letter}, we report for the first time on the three-loop results for $n=5$.
In order to achieve this, we follow two complementary approaches.

The first approach employs the symbol bootstrap method. 
In order to fully determine the answer, we develop novel constraints from a near-collinear expansion, inspired by ref. \cite{Alday:2010ku}. {For the purposes of this {\it Letter}, we perform a preliminary analysis that uses minimal input about the integrable spectrum of flux tube excitations \cite{Basso:2010in}.} 
Developing this approach further, {toward a complete integrability-based description of eq. \eqref{eq:defFintro}}, may unlock powerful new constraints on $F_n$. This could {open up novel opportunities to bootstrap this quantity at higher loop orders, and to perhaps even compute it at finite coupling}. 

The second approach is a first-principles calculation.
We use the latest developments in integral reduction techniques in order to relate the known integrand of $F_5$ at three loops, cf. ref.~\cite{Ambrosio:2013pba}, to the results for Feynman integrals from ref.~\cite{Chicherin:2025}.
This approach allows us to verify the symbol result obtained from the bootstrap method, and moreover yields the full function-level answer. 

In this way, we are able to determine the answer in two completely complementary ways. This is the first physical application of the analytic three-loop five-point Feynman integrals.

This \textit{Letter} is organized as follows. We briefly introduce the Wilson loop with Lagrangian insertion in section \ref{sec:Wilon_loop_Lagrangian_insertion}, and discuss what is known about its leading singularities and the relevant function space. In section~\ref{sec:nearcollinear}, we compute the near-collinear expansion of our observable.
Based on those results, we calculate the three-loop pentagonal Wilson loop with Lagrangian insertion in section \ref{sec:bootstrap}, via the symbol bootstrap method. 
In section \ref{sec:discussion}, we discuss properties of the result.
In section \ref{sec:IBP}, we independently verify our symbol result through a direct calculation based on integral reduction.
Finally, the summary and outlook are given in section~\ref{sec:Summary_Outlook}.

\section{Perturbative structure}
\label{sec:Wilon_loop_Lagrangian_insertion}
    We work in the weak coupling perturbative expansion and in the large $N_c$ limit,
    \begin{equation}
    F_n=\sum_{L\ge0} (g^2)^{L+1}F_n^{(L)}\,, \quad g^2 = \frac{g^2_{\text{YM}}N_c}{16} \,.
    \end{equation}
The Born level results $F^{(0)}_n$ and the one-loop corrections $F^{(1)}_n$ have been computed for arbitrary $n$ in \cite{Chicherin:2022bov}. 
The perturbative results for four and five cusp cases have been obtained up to three and two loops in refs.~\cite{Alday:2012hy,Alday:2013ip,Henn:2019swt} and in~\cite{Chicherin:2022bov,Chicherin:2022zxo}, respectively.
Due to the dual conformal symmetry of sYM, the observable depends on $3n-11$ cross-ratios \cite{Alday:2011ga}, which is the same number of kinematic variables as an $n$ massless particle scattering process in generic theories, such as QCD.
Thanks to the amplitude-Wilson loop duality \cite{Alday:2007hr,Drummond:2007aua,Brandhuber:2007yx} at the level of loop integrands \cite{Mason:2010yk}, the loop integrand of $F^{(L)}_n$ can be constructed from the (integrand of the logarithm of the) ($L+1$)-loop MHV amplitude. The latter is known up to six loops for the five-particle case \cite{Ambrosio:2013pba}.
    
Let us focus on the case $n=5$ from now on.
The $L$-loop contribution to this observable can be written as
\begin{equation} \label{eq:F5_LS}
	F_5^{(L)}=\sum_{i=0}^5 r_i g_i^{(L)} \,,
\end{equation}
with the leading singularities $r_i$ given in ref. \cite{Chicherin:2024hes}.
The $g_i^{(L)}$ are expected to be pure functions of transcendental weight $2L$.
Since in the following we focus on the five-particle case, we will denote $F=F_{5}$ for simpler notation.

Let us discuss in more detail what is known about the function space $g_{i}$ at three loops.
Recently, all three-loop, five-point, planar massless Feynman integrals have been calculated analytically in refs.~\cite{Liu:2024ont,Chicherin:2025}, using the canonical differential equation method \cite{Henn:2013pwa}. 
It turns out that the result for all 
 planar massless Feynman integrals involves $56$ symbol letters, confirming a conjecture made in ref.~\cite{Chicherin:2024hes}.

The results of ref. \cite{ Chicherin:2025} allow us to extract the specific weight-six symbol space needed for describing $F^{(3)}$.
  We organize the symbols according to their transformation properties under the $D_5$ dihedral group. 
In order to achieve this, we use finite field computational tools \cite{Peraro:2019svx,spasm}.
We find $2220$ linearly independent weight-six symbols.
It turns out that the number of symbols required to construct $F^{(3)}$ is slightly different,  
as we explain presently.
$F^{(3)}$ contains both the genuine three-loop
integrals, and products of lower-loop five-point integrals. Taking these into account gives a total of $4729$ symbols.

This number can be reduced by considering the conjectured duality relation with all-plus amplitudes \cite{Chicherin:2022bov}.
This relation constrains the function space, because it relates certain quantities obtained from products to other quantities that are expressed in terms of three-loop integrals only. 
Before using this information in order to write a bootstrap ansatz for eq. (\ref{eq:F5_LS}), let us first discuss constraints that come from a near-collinear expansion of the Wilson loop contour.

\section{Near-collinear expansion}
\label{sec:nearcollinear}

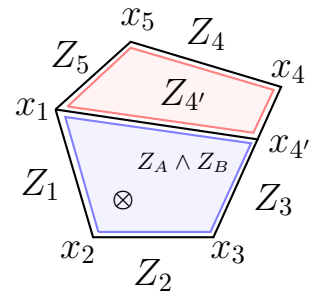
\begin{figure}
    \centering
    
        \begin{tikzpicture}[thick, scale=1]
		\coordinate (2) at (0,   0);
		\coordinate (3) at (1.6,   0);
		\coordinate (4) at (2.5, 2);
		\coordinate (5) at (0.5,   2.6);
		\coordinate (1) at (-0.5,1.7);
		\coordinate (6) at (2.2,1.3);
		\node at (-0.8,1.7) {\Large$x_1$};
		\node at (-0.2,-0.2) {\Large$x_2$};
		\node at (1.8,-0.2) {\Large$x_3$};
		\node at (2.6,2.2) {\Large$x_4$};
		\node at (0.6,2.9) {\Large$x_5$};
		\draw (1)--(2)--(3)--(4)--(5)--(1);
		\draw (1)--(6);
		\coordinate (a) at (0.07,   0.07);
		\coordinate (b) at (-0.37,   1.6);
		\coordinate (c) at (2.1, 1.25);
		\coordinate (d) at (1.56,   0.07);
		\filldraw[fill=blue,draw=blue,fill opacity=0.05,draw opacity=0.5]
			(a)--(b)--(c)--(d)--(a);
		\coordinate (a1) at (-0.34,   1.74);
		\coordinate (b1) at (0.5,   2.52);
		\coordinate (c1) at (2.4, 1.96);
		\coordinate (d1) at (2.13, 1.39);
		\filldraw[fill=red,draw=red,fill opacity=0.05,draw opacity=0.5]
			(a1)--(b1)--(c1)--(d1)--(a1);
		\node at (-0.7,0.7) {\Large$Z_1$};
		\node at (0.8,-0.5) {\Large$Z_2$};
		\node at (2.62,1.2) {\Large$x_{4^\prime}$};
		\node at (2.4,0.5) {\Large$Z_3$};
		\node at (-0.3,2.4) {\Large$Z_5$};
		\node at (1.5,2.7) {\Large$Z_4$};
		\node at (1.2,1.9) {\Large$Z_{4^\prime}$};
		\node at (1.2,1) {\small$Z_A\wedge Z_B$};
		\node at (0.4,0.5) {$\boldsymbol{\otimes}$};
	\end{tikzpicture}
    \caption{
    Quadrilateral and pentagonal Wilson loop contours with cusps $x_i = Z_{i-1} \wedge Z_i$ and $x_{4'} = Z_3 \wedge Z_{4'}$, and the Lagrangian coordinate $x_0 = Z_A \wedge Z_B$. In the collinear limit, the polygon flattens, i.e. $Z_4, Z_5 \to Z_{4'}$, $x_{4} \to x_{4'}$ and $x_5 \in [x_1;x_{4'}]$.}
    \label{fig:OPE}
\end{figure}

The operator product expansion (OPE) for null Wilson loops \cite{Alday:2010ku}, reinforced by integrability of the flux-tube excitations \cite{Basso:2010in,Basso:2013pxa} in planar ${\cal N}=4$ sYM, is a powerful computational tool providing a description of the scattering amplitudes \cite{Basso:2013vsa,Basso:2013aha,Basso:2014koa,Basso:2014hfa,Basso:2015rta,Basso:2015uxa,Belitsky:2014sla,Belitsky:2014lta,Belitsky:2016vyq} and form-factors of half-BPS operators \cite{Sever:2020jjx,Sever:2021nsq,Sever:2021xga,Basso:2023bwv} at finite coupling. 
The multi-loop perturbative predictions of the OPE have been extensively exploited in the bootstrap \cite{Gaiotto:2011dt,Dixon:2011pw,Dixon:2013eka,Dixon:2014voa,Dixon:2014iba,Dixon:2015iva,Caron-Huot:2019vjl,Dixon:2020bbt,Dixon:2022rse,Basso:2024hlx} of these observables at weak coupling. In particular, the OPE controls the expansion of the appropriately regularized Wilson loop (amplitude) around the limit where a pair of its adjacent edges (momenta) becomes collinear. 
The power corrections to this limit, i.e. the near-collinear terms, are controlled by a finite number of excitations of the flux tube.
For the purposes of this {\it Letter}, we will mostly focus on the linear corrections, which originate from single-particle gluon excitations, and which have a particularly simple form.

We employ the flux-tube picture of the null Wilson loop OPE in the new setting of the Wilson loop with Lagrangian insertion, cf. eq. \eqref{eq:defFintro}, to describe the near-collinear expansion. 
We work in the weak coupling regime.
$F_5 \to F_4$ is the lowest possible multiplicity instance of the collinear limit. Thus, $F_4(z)$, which is a function of one cross-ratio $z$,
\begin{align}
z = \frac{x_{13}^2 x_{20}^2 x_{4'0}^2}{x_{24'}^2 x_{10}^2 x_{3¨0}^2} \,, \label{eq:zccr}
\end{align}
is the vacuum configuration, and $F_5$ is a sum of flux-tube excitations propagating on top of this vacuum. 

Momentum twistors are a convenient way to introduce an OPE-friendly parametrization of the kinematics. Thanks to the conformal symmetry, the contour of $F_4$ is the reference square of \cite{Basso:2014koa}, formed by the twistors $Z_1,Z_2,Z_3,Z_{4'}$,
\begin{equation}
\begin{aligned}
& Z_1 = (0,1,0,0) \,,\;
Z_2 = (0,0,0,1) \,,  \\
& Z_3 = (1,0,0,0) \,, \,
Z_{4'} = (0,0,1,0) \,. \label{eq:refquare}
\end{aligned}
\end{equation}
The usual OPE variables \cite{Alday:2010ku,Basso:2013vsa} corresponding to symmetries of the reference square,
\begin{equation}
\begin{aligned}
T = e^{-\tau} \,, \quad
S = e^{\sigma} \,, \quad F= e^{- i \phi}\,, \label{eq:TSF}
\end{aligned}
\end{equation}
define the pentagonal contour of $F_5$ formed by twistors $Z_1,Z_2,Z_3,Z_{4},Z_5$, where \cite{Basso:2013aha},
\begin{equation}
\begin{aligned}
&Z_4 = \biggl( \sqrt{F} S ,0 , \frac{1}{\sqrt{F} T} , \frac{T}{\sqrt{F}} \biggr) , \\
& Z_5 = \biggl( 0 , \frac{\sqrt{F}}{S} , \frac{1}{\sqrt{F}T} , 0 \biggr)\,.
\end{aligned}
\end{equation}
The space-time coordinates $x_i = Z_{i-1} \wedge Z_{i}$ are cusps of the polygon, while the twistors correspond to the light-like edges, cf. Fig.~\ref{fig:OPE}. In the collinear limit $T \to 0$, the pentagon flattens, $x_4 \to x_{4'}$ and $x_5 \in [x_1 ; x_{4'}]$. Finally, we find convenient to parametrize the Lagrangian coordinate by a twistor line, $x_0 = Z_{A} \wedge Z_{B}$,
\begin{align}
Z_{A} = (Z,0,1,-1) \,, \quad
Z_{B} = (1,1,-Z,0) \,.
\end{align}
Here $Z$ satisfies $Z^2 = -1-z$. 
Thus, the kinematics of 
\begin{align}
G= F_5/F_4\,,
\end{align}
is parametrized by the four OPE variables $T,S,F,Z$.

Expanding the ratio $G$ in the near-collinear regime $T \to 0$, and subsequently for small $Z$, based on the OPE picture we expect the following form,
 \begin{equation}
 \begin{aligned}
& G^{(L)} = TF \sum^{L}_{\ell = 0}\sum_{k \geq 0} \log^\ell(T) Z^{1+2k} f_{\ell,k,\bar F}(S)  \\
& + \frac{T}{F} \sum^{L-1}_{\ell = 0}\sum_{k \geq 0} \log^\ell(T) Z^{1+2k} f_{\ell,k, F}(S) + O(T^2) \,. \label{eq:expandG}
\end{aligned}
\end{equation}
A number of comments are in order.
In eq. (\ref{eq:expandG}),
the powers $T^n$ correspond to twist-$n$ excitations with energy $E = n + {\cal{O}}(g^2)$. 
The spectrum of excitations is gapped, so the expansion starts at order $T^1$, which corresponds to a gluon. We do not expect negative or zero powers of $T$.
Powers of $F$ count the $U(1)$ charge (helicity). For example, gluons can lead to $F$ or $1/F$.

Indeed, at Born level we find,
\begin{align}
 G^{(0)} = 1 - T F Z \frac{S}{1+S^2} + O(T^2) \,.  \label{eq:Gborn}
 \end{align}
Note that the asymmetry between $F$-terms and $1/F$-terms, which are absent in eq. (\ref{eq:Gborn}), is due to the $U(1)$ charge of the chiral Lagrangian.
 
Using the available perturbative data for $L=1,2$, we find agreement with eq. (\ref{eq:expandG}). In practice, we expanded up to ${\cal O}(Z^{50})$, and found
that $f_{\ell,k,\bar F}$ and $f_{\ell,k,F}$ are linear combinations of harmonic polylogarithms \cite{Remiddi:1999ew} of argument $S^2$, and  
of transcendental weight $\leq 2L-\ell$ and $\leq 2L-\ell-1 $, respectively, with rational coefficients in $S$.

Let us discuss the OPE picture in more detail, in order to make further quantitative predictions.
We consider single-particle flux tube excitations, and label them by their type. 
In the space-time picture, they correspond to field insertions in the large-spin single-trace operators \cite{Alday:2007mf} which are holographically dual to excitations of the GKP string \cite{Gubser:2002tv}. Their integrable spectrum is parametrized by rapidity $u$. An excitation of type ${\bf a}$ carries energy $E_{\bf a}(u)$, momentum $p_{\bf a}(u)$, and $U(1)$ charge $m_{\bf a}$, corresponding to the conformal symmetries 
of the reference square.  
Their contribution to the ratio $G$ is as follows, 
\begin{align}
G_{\bf a} =
\int\limits^{+\infty}_{-\infty} \frac{du}{2\pi} \widehat{\mu}_{\bf a}(u)\, Q_{\bf a}(u,Z) \,, \label{eq:Ga}
\end{align}
where 
\begin{align}
    \widehat{\mu}_{\bf a}(u) = e^{-E_{\bf a}(u) \tau + i p_{\bf a}(u) \sigma + i \phi m_{\bf a}} \mu_{\bf a}(u) \,,
 \end{align}
    and the exponential factor describes propagation of the type ${\bf a}$ excitation through the reference square, with the measure $\mu_{\bf a}(u)$. The transition form factor $Q_{\bf a}$ is a Taylor series in $Z$, and it describes the transition of the excitation in the presence of the Lagrangian operator. 

The gluon excitations ${\bf a} = F, \bar{F}$ are the simplest to analyse \cite{Gaiotto:2011dt}. They have the lowest energy $E_{F}(u) = 1 + O(g^2)$. Thus, they are the leading contribution of order $T^1$ in the near-collinear limit, cf. eq. \eqref{eq:expandG}. 
They do not mix with the multi-particle excitations.  
Also, $p_F(u) = 2u + O(g^2)$ and $
\mu_F(u) = O(g^2)$ \cite{Basso:2013vsa}.

On the one hand, the spectrum of flux-tube excitations is integrable and their energies, momenta, and measures have been calculated at finite coupling \cite{Basso:2010in}, and they are known at any perturbative order. 
On the other hand, we can glean information on $Q_{\bf a}$ from the known perturbative data on $G^{(L)}$ with $L \leq 2$. 
Comparing the OPE expansion, eq. \eqref{eq:expandG}, with the single-gluon contributions given in eq. \eqref{eq:Ga}, one finds perturbative expansions of $Q_{F}$ and $Q_{\bar{F}}$ up to order $(g^2)^{L-1}$. Using them, we can calculate, from eq. \eqref{eq:Ga}, the logarithmically enhanced contributions $T^1 \log^{\ell'} (T)$ in the expansion at higher loops, i.e. for $G^{(L+\ell)}$ with $\ell\leq \ell' \leq \ell+L$. 

For example, from the Born-level eq.~\eqref{eq:Gborn} we find
\begin{align}
Q_{\bar F}(u,Z) = \frac{Z}{g^2} \left(u^2 + \frac14\right) + O(g^0)\, \label{eq:QFbar}  
\end{align}
that is sufficient to calculate the leading logarithms $T^1 \log^L(T)$ of $G^{(L)}$ at any $L$. The prediction is
\begin{align}
G^{(L)} = -\frac{T F Z}{2 L!} \log^L(T)\int\limits^{+\infty}_{-\infty}\frac{du}{2\pi} \frac{ \left[ E^{(1)}_{\frac{3}{2}}(u)\right]^L e^{2iu \sigma}}{\cosh(\pi u)}\,,
\end{align}
where $E^{(1)}_s(u) = 2\psi(s+ i u) + 2\psi(s-iu)-2\psi(1)$ is the one-loop energy correction for excitations of conformal spin $s$ \cite{Basso:2010in,Belitsky:2006en}, and we omit terms of order ${\cal O}(T \log^{L-1}(T))$. Following this approach, we can calculate all logarithmically enhanced contributions $T^1\log^{\ell}(T)$ with $1\leq \ell \leq 3$ at three loops. For reference, the leading term of the perturbative expansion of the gluon transition is
\begin{equation}
\begin{aligned}\label{eq:QF}
& Q_{F}(u,Z) = Z + O(g^2) \,.
\end{aligned}
\end{equation}
We further conjecture an all-loop relation between the transitions
\begin{align}\label{eq:QFbarQF}
Q_{\bar F}(u,Z) = \frac{1}{g^2}x(u + {\textstyle\frac{i}{2}})x(u - {\textstyle\frac{i}{2}}) Q_{F}(u,Z) \,,
\end{align}
where $x(u)=\frac12
\left(u+\sqrt{u^2-4g^2}\right)$ is the Zhukowsky variable.
We find that eq.~(\ref{eq:QFbarQF}) is compatible with the two-loop data. 

We also considered the OPE contributions from the twist-two ($T^2$) and twist-three ($T^3$) single-particle excitations.
Multi-particle gluon  excitations could mix with them at the level of $(TF)^n \log^1(T)$ and $(T/F)^n \log^0(T)$ OPE terms with twist $n=2,3$ in 
$G^{(3)}$. 
However, higher powers of $\log(T)$ receive single-particle gluon bound state contributions only,
which are under control. 
Finally, we also considered $T^2 F^0$ terms, which correspond to single-particle twist-two excitations of zero $U(1)$ charge \cite{Basso:2013aha}.

\section{Bootstrap}
\label{sec:bootstrap}

\begin{table}[t]
\begin{tabular}{lc}
\toprule
Condition on $F^{(3)}$ & No. of constraints \\
\midrule
collinear limit $p_4|| p_5$ & 682 \\
soft limit $p_5 \to 0$ & 237 \\
no spurious poles & 1693 \\
dependence on $s_{ij}/s_{kl}$ & 778 \\
duality to all-plus & 1182 \\
\midrule {\bf Total} & 2868 \\
\bottomrule
\end{tabular}
\caption{\label{tab:boot} Counting of the bootstrap constraints on the three-loop ansatz for $F^{(3)}$.}
\end{table}

We fix the five-cusp observable at symbol level relying on its physical properties, knowledge about the function space for the relevant Feynman graph topologies, and also conjectures about its analytic properties and the near-collinear OPE.

We construct a weight-six symbol ansatz for $F^{(3)}$, using the symbol space discussed above. 
We then impose the dihedral symmetries on the symbols,
\begin{equation}
    \begin{aligned}
& g_0^{(3)} = \rho \circ g_0^{(3)} = \tau \circ g_0^{(3)} \,, \\
& g_1^{(3)} =  \tau \circ \rho \circ g_1^{(3)} 
\end{aligned} 
\end{equation}
where $\tau$ is a cyclic shift and $\rho$ is a reflection, $\tau \circ p_i = p_{i+1}$ and $\rho \circ p_{i} = p_{6-i}$ at $i=1,\ldots,5$, and $g_{j+1}^{(3)} = \tau^j \circ g^{(3)}_1$ for $j=1,\ldots,4$. We find the dimension of the space for 
$g_0^{(3)}$ and $g_1^{(3)}$ to be $511$ and $2403$, respectively.
Therefore we have a total of $2914$ of unknowns at this stage. 
In order to fix them, we require that following conditions:
\begin{itemize}
    \item In both the collinear limit $p_4 || p_5$, and in the soft limit $p_5 \to 0$, we have $F_5 \to F_4$. 
    \item The alphabet letters are organized in dimensionless ratios in the symbols.
    \item The spurious pole of $r_1$ at $s_{25} \to 0$ is suppressed by the vanishing of the accompanying symbol $g^{(3)}_1$.
    \item We assume that the duality to all-plus amplitudes holds. 
    The latter yields a homogeneous constraint that merely requires the all-plus amplitude to lie within the space of three-loop Feynman integrals, without needing its explicit form. 
   (Interestingly, and in contrast, once our bootstrap is successfully completed, it provides a concrete prediction for the all-plus four-loop amplitude.)
\end{itemize}
       In this way, we obtain $2868$ independent  constraints on the $2914$ unknowns in our ansatz, see Table~\ref{tab:boot}.
       
In order to fix the remaining $46$ unknowns, we invoke the OPE description of the near-collinear expansion of $F^{(3)}$ at $p_4 || p_5$, as discussed in section \ref{sec:nearcollinear}.
{The next-to-leading logarithmic OPE terms  $(T/F)\log^1(T)$ in eq. \eqref{eq:expandG} at three loops 
fix 39 unknowns.
All other single-particle contributions from gluons and their bounds states do not provide independent constraints.
However, they serve as valuable consistency checks of the bootstrap approach.
Finally, we fix the remaining 5 unknowns by considering the leading logarithmic OPE term $T^2 F^0 \log^3(T)$ for the effective single-particle twist-two excitation $\psi \bar{\psi}$ of zero $U(1)$ charge.}

In summary, combining knowledge of the function space, restrictions from the conjectured duality to all-plus amplitudes, physical constraints, and near-collinear conditions, we are able to uniquely fix the bootstrap ansatz. 
We provide the symbol results for $F^{(L)}$ at $L=3$ (as well as the lower loop ones, for completeness) in the following repository,
\href{https://uva-hva.gitlab.host/universeplus/pentagon-wilson-loop-with-lagrangian-insertion-at-three-loops}{https://uva-hva.gitlab.host/universeplus}.

\section{Discussion of the result}
\label{sec:discussion}
$F^{(3)}$ is finite and has uniform weight six. 
Although generic planar three-loop five-point Feynman integrals contain $56$ alphabet letters, $F^{(3)}$ contains $30$ letters only, namely $25$ two-loop five-point planar letters, 
and five additional three-loop letters
that had been predicted in ref. \cite{Chicherin:2024hes}.
For convenience and to make this {\it{Letter}} self-contained, we list the relevant letters in the Supplemental Material.

It is interesting to study how the symbol letters appear in the symbol of $F^{(3)}$, and to look for patterns across loop orders.
    We find that the letters appearing in the first three entries coincide with the first three entries appearing at two loops. The novel three-loop letters appear in the fourth entry only.    
    The last entries of $g_0^{(3)}$ depend on $19$ linear combinations of the letters $w_1, \ldots w_5, w_{11},\ldots w_{25}$ only. The last entries of $g_1^{(3)}$ depend on $15$ linear combinations of the letters $w_{1,...,5}$, $w_8$, $w_{10}$, $w_{11,...,20}$, $w_{22}$, $w_{24}$ only. 

Let us also comment on limits of the answer. $F_{5}$ depends on five Mandelstam variables (via four dimensionless ratios), so there are many interesting configurations one can consider. Many of these have already been explored in the bootstrap construction.
However, one further interesting limit to consider is the multi-Regge limit.
This was done in section 4.5. of ref. \cite{Chicherin:2022zxo} for the two-loop case, where it was found that the symbol of 
$\log \bigl( F/F^{(0)} \bigr)$ vanishes in that limit, up to two loops. (Note, however, that this relation is not exact, due to beyond-the-symbol terms.)
We verify that the same relation holds also at order $g^6$.

\section{Direct verification from integral reduction}\label{sec:IBP}

In this section, we describe a complementary approach to obtaining the three-loop five-point pentagonal Wilson loop with Lagrangian insertion, based on explicit loop integration. This serves as an independent check of the bootstrap construction discussed in section \ref{sec:bootstrap}.

A generic Feynman integral contributing to the Wilson loop can be written as a graph in dual space.  
The latter corresponds either to products of lower loop graphs (such as cubic product of pentagons, or a product of a one-loop pentagon and a two-loop pentabox integrals), which are known from references \cite{Gehrmann:2015bfy,Gehrmann:2018yef,Chicherin:2020oor}, or to genuine three-loop integrals computed in ref. \cite{Chicherin:2025} (or permutations thereof).
 The reductions are performed using the new \verb|SpanningCut| feature in the latest version of the package \texttt{NeatIBP}~\cite{Wu:2023upw,Wu:2024paw,Wu:2025aeg}.

Although $F^{(3)}$ is finite in four dimensions, each individual Feynman integral develops divergences upon loop integration, of up to order $\epsilon^{-6}$. Therefore the cancellation of these divergences is an important check of our calculation.
 Taking the symbol of the final result, we find perfect agreement with the outcome of the symbol bootstrap approach.

\section{Summary and Outlook}\label{sec:sum}

In this \textit{Letter}, we computed the pentagonal Wilson loop with a Lagrangian insertion in planar $\mathcal{N}=4$ sYM theory at three-loop order by employing a bootstrap approach.
An essential new input came from predictions we derived by analyzing the expected structure in a near-collinear expansion. 
We independently verified our result by directly reducing the integrand onto the canonical basis from ref. \cite{Chicherin:2025}. 
This means that we also have access to numerical evaluation.
Our result constitutes the first physical observable in the newly uncovered three-loop five-point planar function space, thereby pushing the frontier of QCD perturbative computations.

We anticipate that our results can be extended to function level (i.e., going beyond the proof-of-principle numerical evaluation from ref. \cite{Chicherin:2025}), which will be extremely interesting in the perspective of testing the conjectured positivity~\cite{Chicherin:2022zxo,Chicherin:2024hes} and complete monotonicity properties ~\cite{Henn:2024qwe}.  Furthermore, via the conjecture of ref.~\cite{Chicherin:2022bov}, our result provides a prediction for the maximally transcendental part of the planar four-loop five-point all-plus amplitude in pure Yang-Mills theory. 

Finally, it would be very interesting to develop further the near-collinear expansion, or OPE approach. We used this for the first time in the context of Wilson loops with Lagrangian insertions. Although the predictions gleaned from this were sufficient to complete the three-loop bootstrap ansatz, potentially much stronger constraints could be derived. 
This may ultimately even give access to finite coupling results for the observables studied here.

\section{Acknowledgements}
\label{sec:Summary_Outlook}

We are indebted to Alexander Tumanov for initial collaboration on this project. We thank Benjamin Basso, James Drummond, Ömer Gurdoğan,  and Qinglin Yang for discussions.
This work is supported by the European Union (ERC, UNIVERSE PLUS, 101118787).
Views and opinions expressed are however those of the
authors only and do not necessarily reflect those of the
European Union or the European Research Council Executive Agency. Neither the European Union nor the
granting authority can be held responsible for them. DC
is supported by ANR-24-CE31-7996. YZ is supported
by NSFC through Grant No. 12575078 and 12247103.

\bibliographystyle{apsrev4-1}
\bibliography{main}

\begin{thebibliography}{62}%
\makeatletter
\providecommand \@ifxundefined [1]{%
 \@ifx{#1\undefined}
}%
\providecommand \@ifnum [1]{%
 \ifnum #1\expandafter \@firstoftwo
 \else \expandafter \@secondoftwo
 \fi
}%
\providecommand \@ifx [1]{%
 \ifx #1\expandafter \@firstoftwo
 \else \expandafter \@secondoftwo
 \fi
}%
\providecommand \natexlab [1]{#1}%
\providecommand \enquote  [1]{``#1''}%
\providecommand \bibnamefont  [1]{#1}%
\providecommand \bibfnamefont [1]{#1}%
\providecommand \citenamefont [1]{#1}%
\providecommand \href@noop [0]{\@secondoftwo}%
\providecommand \href [0]{\begingroup \@sanitize@url \@href}%
\providecommand \@href[1]{\@@startlink{#1}\@@href}%
\providecommand \@@href[1]{\endgroup#1\@@endlink}%
\providecommand \@sanitize@url [0]{\catcode `\\12\catcode `\$12\catcode `\&12\catcode `\#12\catcode `\^12\catcode `\_12\catcode `\%12\relax}%
\providecommand \@@startlink[1]{}%
\providecommand \@@endlink[0]{}%
\providecommand \url  [0]{\begingroup\@sanitize@url \@url }%
\providecommand \@url [1]{\endgroup\@href {#1}{\urlprefix }}%
\providecommand \urlprefix  [0]{URL }%
\providecommand \Eprint [0]{\href }%
\providecommand \doibase [0]{http://dx.doi.org/}%
\providecommand \selectlanguage [0]{\@gobble}%
\providecommand \bibinfo  [0]{\@secondoftwo}%
\providecommand \bibfield  [0]{\@secondoftwo}%
\providecommand \translation [1]{[#1]}%
\providecommand \BibitemOpen [0]{}%
\providecommand \bibitemStop [0]{}%
\providecommand \bibitemNoStop [0]{.\EOS\space}%
\providecommand \EOS [0]{\spacefactor3000\relax}%
\providecommand \BibitemShut  [1]{\csname bibitem#1\endcsname}%
\let\auto@bib@innerbib\@empty
\bibitem [{\citenamefont {Henn}(2021)}]{Henn:2020omi}%
  \BibitemOpen
  \bibfield  {author} {\bibinfo {author} {\bibfnamefont {J.~M.}\ \bibnamefont {Henn}},\ }\href {\doibase 10.1146/annurev-nucl-102819-100428} {\bibfield  {journal} {\bibinfo  {journal} {Ann. Rev. Nucl. Part. Sci.}\ }\textbf {\bibinfo {volume} {71}},\ \bibinfo {pages} {87} (\bibinfo {year} {2021})},\ \Eprint {http://arxiv.org/abs/2006.00361} {arXiv:2006.00361 [hep-th]} \BibitemShut {NoStop}%
\bibitem [{\citenamefont {Arkani-Hamed}\ \emph {et~al.}(2022)\citenamefont {Arkani-Hamed}, \citenamefont {Dixon}, \citenamefont {McLeod}, \citenamefont {Spradlin}, \citenamefont {Trnka},\ and\ \citenamefont {Volovich}}]{Arkani-Hamed:2022rwr}%
  \BibitemOpen
  \bibfield  {author} {\bibinfo {author} {\bibfnamefont {N.}~\bibnamefont {Arkani-Hamed}}, \bibinfo {author} {\bibfnamefont {L.~J.}\ \bibnamefont {Dixon}}, \bibinfo {author} {\bibfnamefont {A.~J.}\ \bibnamefont {McLeod}}, \bibinfo {author} {\bibfnamefont {M.}~\bibnamefont {Spradlin}}, \bibinfo {author} {\bibfnamefont {J.}~\bibnamefont {Trnka}}, \ and\ \bibinfo {author} {\bibfnamefont {A.}~\bibnamefont {Volovich}},\ }\href@noop {} {\  (\bibinfo {year} {2022})},\ \Eprint {http://arxiv.org/abs/2207.10636} {arXiv:2207.10636 [hep-th]} \BibitemShut {NoStop}%
\bibitem [{\citenamefont {Chicherin}\ and\ \citenamefont {Henn}(2022{\natexlab{a}})}]{Chicherin:2022bov}%
  \BibitemOpen
  \bibfield  {author} {\bibinfo {author} {\bibfnamefont {D.}~\bibnamefont {Chicherin}}\ and\ \bibinfo {author} {\bibfnamefont {J.~M.}\ \bibnamefont {Henn}},\ }\href {\doibase 10.1007/JHEP07(2022)057} {\bibfield  {journal} {\bibinfo  {journal} {JHEP}\ }\textbf {\bibinfo {volume} {07}},\ \bibinfo {pages} {057} (\bibinfo {year} {2022}{\natexlab{a}})},\ \Eprint {http://arxiv.org/abs/2202.05596} {arXiv:2202.05596 [hep-th]} \BibitemShut {NoStop}%
\bibitem [{\citenamefont {Arkani-Hamed}\ and\ \citenamefont {Trnka}(2014)}]{Arkani-Hamed:2013jha}%
  \BibitemOpen
  \bibfield  {author} {\bibinfo {author} {\bibfnamefont {N.}~\bibnamefont {Arkani-Hamed}}\ and\ \bibinfo {author} {\bibfnamefont {J.}~\bibnamefont {Trnka}},\ }\href {\doibase 10.1007/JHEP10(2014)030} {\bibfield  {journal} {\bibinfo  {journal} {JHEP}\ }\textbf {\bibinfo {volume} {10}},\ \bibinfo {pages} {030} (\bibinfo {year} {2014})},\ \Eprint {http://arxiv.org/abs/1312.2007} {arXiv:1312.2007 [hep-th]} \BibitemShut {NoStop}%
\bibitem [{\citenamefont {Chicherin}\ and\ \citenamefont {Henn}(2022{\natexlab{b}})}]{Chicherin:2022zxo}%
  \BibitemOpen
  \bibfield  {author} {\bibinfo {author} {\bibfnamefont {D.}~\bibnamefont {Chicherin}}\ and\ \bibinfo {author} {\bibfnamefont {J.}~\bibnamefont {Henn}},\ }\href {\doibase 10.1007/JHEP07(2022)038} {\bibfield  {journal} {\bibinfo  {journal} {JHEP}\ }\textbf {\bibinfo {volume} {07}},\ \bibinfo {pages} {038} (\bibinfo {year} {2022}{\natexlab{b}})},\ \Eprint {http://arxiv.org/abs/2204.00329} {arXiv:2204.00329 [hep-th]} \BibitemShut {NoStop}%
\bibitem [{\citenamefont {Brown}\ \emph {et~al.}(2025)\citenamefont {Brown}, \citenamefont {Henn}, \citenamefont {Mazzucchelli},\ and\ \citenamefont {Trnka}}]{Brown:2025plq}%
  \BibitemOpen
  \bibfield  {author} {\bibinfo {author} {\bibfnamefont {T.~V.}\ \bibnamefont {Brown}}, \bibinfo {author} {\bibfnamefont {J.~M.}\ \bibnamefont {Henn}}, \bibinfo {author} {\bibfnamefont {E.}~\bibnamefont {Mazzucchelli}}, \ and\ \bibinfo {author} {\bibfnamefont {J.}~\bibnamefont {Trnka}},\ }\href@noop {} {\  (\bibinfo {year} {2025})},\ \Eprint {http://arxiv.org/abs/2503.17185} {arXiv:2503.17185 [hep-th]} \BibitemShut {NoStop}%
\bibitem [{\citenamefont {Carr{\^o}lo}\ \emph {et~al.}(2025)\citenamefont {Carr{\^o}lo}, \citenamefont {Chicherin}, \citenamefont {Henn}, \citenamefont {Yang},\ and\ \citenamefont {Zhang}}]{Carrolo:2025pue}%
  \BibitemOpen
  \bibfield  {author} {\bibinfo {author} {\bibfnamefont {S.}~\bibnamefont {Carr{\^o}lo}}, \bibinfo {author} {\bibfnamefont {D.}~\bibnamefont {Chicherin}}, \bibinfo {author} {\bibfnamefont {J.}~\bibnamefont {Henn}}, \bibinfo {author} {\bibfnamefont {Q.}~\bibnamefont {Yang}}, \ and\ \bibinfo {author} {\bibfnamefont {Y.}~\bibnamefont {Zhang}},\ }\href {\doibase 10.1007/JHEP07(2025)214} {\bibfield  {journal} {\bibinfo  {journal} {JHEP}\ }\textbf {\bibinfo {volume} {07}},\ \bibinfo {pages} {214} (\bibinfo {year} {2025})},\ \Eprint {http://arxiv.org/abs/2505.01245} {arXiv:2505.01245 [hep-th]} \BibitemShut {NoStop}%
\bibitem [{\citenamefont {Abreu}\ \emph {et~al.}(2024)\citenamefont {Abreu}, \citenamefont {Monni}, \citenamefont {Page},\ and\ \citenamefont {Usovitsch}}]{Abreu:2024fei}%
  \BibitemOpen
  \bibfield  {author} {\bibinfo {author} {\bibfnamefont {S.}~\bibnamefont {Abreu}}, \bibinfo {author} {\bibfnamefont {P.~F.}\ \bibnamefont {Monni}}, \bibinfo {author} {\bibfnamefont {B.}~\bibnamefont {Page}}, \ and\ \bibinfo {author} {\bibfnamefont {J.}~\bibnamefont {Usovitsch}},\ }\href@noop {} {\  (\bibinfo {year} {2024})},\ \Eprint {http://arxiv.org/abs/2412.19884} {arXiv:2412.19884 [hep-ph]} \BibitemShut {NoStop}%
\bibitem [{\citenamefont {Henn}\ \emph {et~al.}(2025)\citenamefont {Henn}, \citenamefont {Matija{\v{s}}i{\'c}}, \citenamefont {Miczajka}, \citenamefont {Peraro}, \citenamefont {Xu},\ and\ \citenamefont {Zhang}}]{Henn:2025xrc}%
  \BibitemOpen
  \bibfield  {author} {\bibinfo {author} {\bibfnamefont {J.}~\bibnamefont {Henn}}, \bibinfo {author} {\bibfnamefont {A.}~\bibnamefont {Matija{\v{s}}i{\'c}}}, \bibinfo {author} {\bibfnamefont {J.}~\bibnamefont {Miczajka}}, \bibinfo {author} {\bibfnamefont {T.}~\bibnamefont {Peraro}}, \bibinfo {author} {\bibfnamefont {Y.}~\bibnamefont {Xu}}, \ and\ \bibinfo {author} {\bibfnamefont {Y.}~\bibnamefont {Zhang}},\ }\href {\doibase 10.1103/zhzd-tj9p} {\bibfield  {journal} {\bibinfo  {journal} {Phys. Rev. Lett.}\ }\textbf {\bibinfo {volume} {135}},\ \bibinfo {pages} {031601} (\bibinfo {year} {2025})},\ \Eprint {http://arxiv.org/abs/2501.01847} {arXiv:2501.01847 [hep-ph]} \BibitemShut {NoStop}%
\bibitem [{\citenamefont {Chicherin}\ \emph {et~al.}(2024)\citenamefont {Chicherin}, \citenamefont {Henn}, \citenamefont {Trnka},\ and\ \citenamefont {Zhang}}]{Chicherin:2024hes}%
  \BibitemOpen
  \bibfield  {author} {\bibinfo {author} {\bibfnamefont {D.}~\bibnamefont {Chicherin}}, \bibinfo {author} {\bibfnamefont {J.}~\bibnamefont {Henn}}, \bibinfo {author} {\bibfnamefont {J.}~\bibnamefont {Trnka}}, \ and\ \bibinfo {author} {\bibfnamefont {S.-Q.}\ \bibnamefont {Zhang}},\ }\href@noop {} {\  (\bibinfo {year} {2024})},\ \Eprint {http://arxiv.org/abs/2410.11456} {arXiv:2410.11456 [hep-th]} \BibitemShut {NoStop}%
\bibitem [{\citenamefont {Chicherin}\ \emph {et~al.}()\citenamefont {Chicherin}, \citenamefont {Wu}, \citenamefont {Wu}, \citenamefont {Xu}, \citenamefont {Zhang},\ and\ \citenamefont {Zhang}}]{Chicherin:2025}%
  \BibitemOpen
  \bibfield  {author} {\bibinfo {author} {\bibfnamefont {D.}~\bibnamefont {Chicherin}}, \bibinfo {author} {\bibfnamefont {Y.}~\bibnamefont {Wu}}, \bibinfo {author} {\bibfnamefont {Z.}~\bibnamefont {Wu}}, \bibinfo {author} {\bibfnamefont {Y.}~\bibnamefont {Xu}}, \bibinfo {author} {\bibfnamefont {S.}~\bibnamefont {Zhang}}, \ and\ \bibinfo {author} {\bibfnamefont {Y.}~\bibnamefont {Zhang}},\ }\href@noop {} {\bibinfo  {journal} {to appear on arXiv}\ }\BibitemShut {NoStop}%
\bibitem [{\citenamefont {Alday}\ \emph {et~al.}(2011{\natexlab{a}})\citenamefont {Alday}, \citenamefont {Gaiotto}, \citenamefont {Maldacena}, \citenamefont {Sever},\ and\ \citenamefont {Vieira}}]{Alday:2010ku}%
  \BibitemOpen
\bibfield  {journal} {  }\bibfield  {author} {\bibinfo {author} {\bibfnamefont {L.~F.}\ \bibnamefont {Alday}}, \bibinfo {author} {\bibfnamefont {D.}~\bibnamefont {Gaiotto}}, \bibinfo {author} {\bibfnamefont {J.}~\bibnamefont {Maldacena}}, \bibinfo {author} {\bibfnamefont {A.}~\bibnamefont {Sever}}, \ and\ \bibinfo {author} {\bibfnamefont {P.}~\bibnamefont {Vieira}},\ }\href {\doibase 10.1007/JHEP04(2011)088} {\bibfield  {journal} {\bibinfo  {journal} {JHEP}\ }\textbf {\bibinfo {volume} {04}},\ \bibinfo {pages} {088} (\bibinfo {year} {2011}{\natexlab{a}})},\ \Eprint {http://arxiv.org/abs/1006.2788} {arXiv:1006.2788 [hep-th]} \BibitemShut {NoStop}%
\bibitem [{\citenamefont {Basso}(2012)}]{Basso:2010in}%
  \BibitemOpen
  \bibfield  {author} {\bibinfo {author} {\bibfnamefont {B.}~\bibnamefont {Basso}},\ }\href {\doibase 10.1016/j.nuclphysb.2011.12.010} {\bibfield  {journal} {\bibinfo  {journal} {Nucl. Phys. B}\ }\textbf {\bibinfo {volume} {857}},\ \bibinfo {pages} {254} (\bibinfo {year} {2012})},\ \Eprint {http://arxiv.org/abs/1010.5237} {arXiv:1010.5237 [hep-th]} \BibitemShut {NoStop}%
\bibitem [{\citenamefont {Ambrosio}\ \emph {et~al.}(2015)\citenamefont {Ambrosio}, \citenamefont {Eden}, \citenamefont {Goddard}, \citenamefont {Heslop},\ and\ \citenamefont {Taylor}}]{Ambrosio:2013pba}%
  \BibitemOpen
  \bibfield  {author} {\bibinfo {author} {\bibfnamefont {R.~G.}\ \bibnamefont {Ambrosio}}, \bibinfo {author} {\bibfnamefont {B.}~\bibnamefont {Eden}}, \bibinfo {author} {\bibfnamefont {T.}~\bibnamefont {Goddard}}, \bibinfo {author} {\bibfnamefont {P.}~\bibnamefont {Heslop}}, \ and\ \bibinfo {author} {\bibfnamefont {C.}~\bibnamefont {Taylor}},\ }\href {\doibase 10.1007/JHEP01(2015)116} {\bibfield  {journal} {\bibinfo  {journal} {JHEP}\ }\textbf {\bibinfo {volume} {01}},\ \bibinfo {pages} {116} (\bibinfo {year} {2015})},\ \Eprint {http://arxiv.org/abs/1312.1163} {arXiv:1312.1163 [hep-th]} \BibitemShut {NoStop}%
\bibitem [{\citenamefont {Alday}\ \emph {et~al.}(2013{\natexlab{a}})\citenamefont {Alday}, \citenamefont {Heslop},\ and\ \citenamefont {Sikorowski}}]{Alday:2012hy}%
  \BibitemOpen
  \bibfield  {author} {\bibinfo {author} {\bibfnamefont {L.~F.}\ \bibnamefont {Alday}}, \bibinfo {author} {\bibfnamefont {P.}~\bibnamefont {Heslop}}, \ and\ \bibinfo {author} {\bibfnamefont {J.}~\bibnamefont {Sikorowski}},\ }\href {\doibase 10.1007/JHEP03(2013)074} {\bibfield  {journal} {\bibinfo  {journal} {JHEP}\ }\textbf {\bibinfo {volume} {03}},\ \bibinfo {pages} {074} (\bibinfo {year} {2013}{\natexlab{a}})},\ \Eprint {http://arxiv.org/abs/1207.4316} {arXiv:1207.4316 [hep-th]} \BibitemShut {NoStop}%
\bibitem [{\citenamefont {Alday}\ \emph {et~al.}(2013{\natexlab{b}})\citenamefont {Alday}, \citenamefont {Henn},\ and\ \citenamefont {Sikorowski}}]{Alday:2013ip}%
  \BibitemOpen
  \bibfield  {author} {\bibinfo {author} {\bibfnamefont {L.~F.}\ \bibnamefont {Alday}}, \bibinfo {author} {\bibfnamefont {J.~M.}\ \bibnamefont {Henn}}, \ and\ \bibinfo {author} {\bibfnamefont {J.}~\bibnamefont {Sikorowski}},\ }\href {\doibase 10.1007/JHEP03(2013)058} {\bibfield  {journal} {\bibinfo  {journal} {JHEP}\ }\textbf {\bibinfo {volume} {03}},\ \bibinfo {pages} {058} (\bibinfo {year} {2013}{\natexlab{b}})},\ \Eprint {http://arxiv.org/abs/1301.0149} {arXiv:1301.0149 [hep-th]} \BibitemShut {NoStop}%
\bibitem [{\citenamefont {Henn}\ \emph {et~al.}(2020)\citenamefont {Henn}, \citenamefont {Korchemsky},\ and\ \citenamefont {Mistlberger}}]{Henn:2019swt}%
  \BibitemOpen
  \bibfield  {author} {\bibinfo {author} {\bibfnamefont {J.~M.}\ \bibnamefont {Henn}}, \bibinfo {author} {\bibfnamefont {G.~P.}\ \bibnamefont {Korchemsky}}, \ and\ \bibinfo {author} {\bibfnamefont {B.}~\bibnamefont {Mistlberger}},\ }\href {\doibase 10.1007/JHEP04(2020)018} {\bibfield  {journal} {\bibinfo  {journal} {JHEP}\ }\textbf {\bibinfo {volume} {04}},\ \bibinfo {pages} {018} (\bibinfo {year} {2020})},\ \Eprint {http://arxiv.org/abs/1911.10174} {arXiv:1911.10174 [hep-th]} \BibitemShut {NoStop}%
\bibitem [{\citenamefont {Alday}\ \emph {et~al.}(2011{\natexlab{b}})\citenamefont {Alday}, \citenamefont {Buchbinder},\ and\ \citenamefont {Tseytlin}}]{Alday:2011ga}%
  \BibitemOpen
  \bibfield  {author} {\bibinfo {author} {\bibfnamefont {L.~F.}\ \bibnamefont {Alday}}, \bibinfo {author} {\bibfnamefont {E.~I.}\ \bibnamefont {Buchbinder}}, \ and\ \bibinfo {author} {\bibfnamefont {A.~A.}\ \bibnamefont {Tseytlin}},\ }\href {\doibase 10.1007/JHEP09(2011)034} {\bibfield  {journal} {\bibinfo  {journal} {JHEP}\ }\textbf {\bibinfo {volume} {09}},\ \bibinfo {pages} {034} (\bibinfo {year} {2011}{\natexlab{b}})},\ \Eprint {http://arxiv.org/abs/1107.5702} {arXiv:1107.5702 [hep-th]} \BibitemShut {NoStop}%
\bibitem [{\citenamefont {Alday}\ and\ \citenamefont {Maldacena}(2007{\natexlab{a}})}]{Alday:2007hr}%
  \BibitemOpen
  \bibfield  {author} {\bibinfo {author} {\bibfnamefont {L.~F.}\ \bibnamefont {Alday}}\ and\ \bibinfo {author} {\bibfnamefont {J.~M.}\ \bibnamefont {Maldacena}},\ }\href {\doibase 10.1088/1126-6708/2007/06/064} {\bibfield  {journal} {\bibinfo  {journal} {JHEP}\ }\textbf {\bibinfo {volume} {06}},\ \bibinfo {pages} {064} (\bibinfo {year} {2007}{\natexlab{a}})},\ \Eprint {http://arxiv.org/abs/0705.0303} {arXiv:0705.0303 [hep-th]} \BibitemShut {NoStop}%
\bibitem [{\citenamefont {Drummond}\ \emph {et~al.}(2008)\citenamefont {Drummond}, \citenamefont {Korchemsky},\ and\ \citenamefont {Sokatchev}}]{Drummond:2007aua}%
  \BibitemOpen
  \bibfield  {author} {\bibinfo {author} {\bibfnamefont {J.~M.}\ \bibnamefont {Drummond}}, \bibinfo {author} {\bibfnamefont {G.~P.}\ \bibnamefont {Korchemsky}}, \ and\ \bibinfo {author} {\bibfnamefont {E.}~\bibnamefont {Sokatchev}},\ }\href {\doibase 10.1016/j.nuclphysb.2007.11.041} {\bibfield  {journal} {\bibinfo  {journal} {Nucl. Phys. B}\ }\textbf {\bibinfo {volume} {795}},\ \bibinfo {pages} {385} (\bibinfo {year} {2008})},\ \Eprint {http://arxiv.org/abs/0707.0243} {arXiv:0707.0243 [hep-th]} \BibitemShut {NoStop}%
\bibitem [{\citenamefont {Brandhuber}\ \emph {et~al.}(2008)\citenamefont {Brandhuber}, \citenamefont {Heslop},\ and\ \citenamefont {Travaglini}}]{Brandhuber:2007yx}%
  \BibitemOpen
  \bibfield  {author} {\bibinfo {author} {\bibfnamefont {A.}~\bibnamefont {Brandhuber}}, \bibinfo {author} {\bibfnamefont {P.}~\bibnamefont {Heslop}}, \ and\ \bibinfo {author} {\bibfnamefont {G.}~\bibnamefont {Travaglini}},\ }\href {\doibase 10.1016/j.nuclphysb.2007.11.002} {\bibfield  {journal} {\bibinfo  {journal} {Nucl. Phys. B}\ }\textbf {\bibinfo {volume} {794}},\ \bibinfo {pages} {231} (\bibinfo {year} {2008})},\ \Eprint {http://arxiv.org/abs/0707.1153} {arXiv:0707.1153 [hep-th]} \BibitemShut {NoStop}%
\bibitem [{\citenamefont {Mason}\ and\ \citenamefont {Skinner}(2010)}]{Mason:2010yk}%
  \BibitemOpen
  \bibfield  {author} {\bibinfo {author} {\bibfnamefont {L.~J.}\ \bibnamefont {Mason}}\ and\ \bibinfo {author} {\bibfnamefont {D.}~\bibnamefont {Skinner}},\ }\href {\doibase 10.1007/JHEP12(2010)018} {\bibfield  {journal} {\bibinfo  {journal} {JHEP}\ }\textbf {\bibinfo {volume} {12}},\ \bibinfo {pages} {018} (\bibinfo {year} {2010})},\ \Eprint {http://arxiv.org/abs/1009.2225} {arXiv:1009.2225 [hep-th]} \BibitemShut {NoStop}%
\bibitem [{\citenamefont {Liu}\ \emph {et~al.}(2025)\citenamefont {Liu}, \citenamefont {Matija{\v{s}}i{\'c}}, \citenamefont {Miczajka}, \citenamefont {Xu}, \citenamefont {Xu},\ and\ \citenamefont {Zhang}}]{Liu:2024ont}%
  \BibitemOpen
  \bibfield  {author} {\bibinfo {author} {\bibfnamefont {Y.}~\bibnamefont {Liu}}, \bibinfo {author} {\bibfnamefont {A.}~\bibnamefont {Matija{\v{s}}i{\'c}}}, \bibinfo {author} {\bibfnamefont {J.}~\bibnamefont {Miczajka}}, \bibinfo {author} {\bibfnamefont {Y.}~\bibnamefont {Xu}}, \bibinfo {author} {\bibfnamefont {Y.}~\bibnamefont {Xu}}, \ and\ \bibinfo {author} {\bibfnamefont {Y.}~\bibnamefont {Zhang}},\ }\href {\doibase 10.1103/qrk2-cym5} {\bibfield  {journal} {\bibinfo  {journal} {Phys. Rev. D}\ }\textbf {\bibinfo {volume} {112}},\ \bibinfo {pages} {016021} (\bibinfo {year} {2025})},\ \Eprint {http://arxiv.org/abs/2411.18697} {arXiv:2411.18697 [hep-ph]} \BibitemShut {NoStop}%
\bibitem [{\citenamefont {Henn}(2013)}]{Henn:2013pwa}%
  \BibitemOpen
  \bibfield  {author} {\bibinfo {author} {\bibfnamefont {J.~M.}\ \bibnamefont {Henn}},\ }\href {\doibase 10.1103/PhysRevLett.110.251601} {\bibfield  {journal} {\bibinfo  {journal} {Phys. Rev. Lett.}\ }\textbf {\bibinfo {volume} {110}},\ \bibinfo {pages} {251601} (\bibinfo {year} {2013})},\ \Eprint {http://arxiv.org/abs/1304.1806} {arXiv:1304.1806 [hep-th]} \BibitemShut {NoStop}%
\bibitem [{\citenamefont {Peraro}(2019)}]{Peraro:2019svx}%
  \BibitemOpen
  \bibfield  {author} {\bibinfo {author} {\bibfnamefont {T.}~\bibnamefont {Peraro}},\ }\href {\doibase 10.1007/JHEP07(2019)031} {\bibfield  {journal} {\bibinfo  {journal} {JHEP}\ }\textbf {\bibinfo {volume} {07}},\ \bibinfo {pages} {031} (\bibinfo {year} {2019})},\ \Eprint {http://arxiv.org/abs/1905.08019} {arXiv:1905.08019 [hep-ph]} \BibitemShut {NoStop}%
\bibitem [{\citenamefont {Bouillaguet}(2023)}]{spasm}%
  \BibitemOpen
  \bibfield  {author} {\bibinfo {author} {\bibfnamefont {C.}~\bibnamefont {Bouillaguet}},\ }\href@noop {} {\emph {\bibinfo {title} {{SpaSM}: a Sparse direct Solver Modulo $p$}}},\ \bibinfo {edition} {v1.3}\ ed. (\bibinfo {year} {2023}),\ \bibinfo {note} {\url{http://github.com/cbouilla/spasm}}\BibitemShut {NoStop}%
\bibitem [{\citenamefont {Basso}\ and\ \citenamefont {Rej}(2014)}]{Basso:2013pxa}%
  \BibitemOpen
  \bibfield  {author} {\bibinfo {author} {\bibfnamefont {B.}~\bibnamefont {Basso}}\ and\ \bibinfo {author} {\bibfnamefont {A.}~\bibnamefont {Rej}},\ }\href {\doibase 10.1016/j.nuclphysb.2013.11.010} {\bibfield  {journal} {\bibinfo  {journal} {Nucl. Phys. B}\ }\textbf {\bibinfo {volume} {879}},\ \bibinfo {pages} {162} (\bibinfo {year} {2014})},\ \Eprint {http://arxiv.org/abs/1306.1741} {arXiv:1306.1741 [hep-th]} \BibitemShut {NoStop}%
\bibitem [{\citenamefont {Basso}\ \emph {et~al.}(2013)\citenamefont {Basso}, \citenamefont {Sever},\ and\ \citenamefont {Vieira}}]{Basso:2013vsa}%
  \BibitemOpen
  \bibfield  {author} {\bibinfo {author} {\bibfnamefont {B.}~\bibnamefont {Basso}}, \bibinfo {author} {\bibfnamefont {A.}~\bibnamefont {Sever}}, \ and\ \bibinfo {author} {\bibfnamefont {P.}~\bibnamefont {Vieira}},\ }\href {\doibase 10.1103/PhysRevLett.111.091602} {\bibfield  {journal} {\bibinfo  {journal} {Phys. Rev. Lett.}\ }\textbf {\bibinfo {volume} {111}},\ \bibinfo {pages} {091602} (\bibinfo {year} {2013})},\ \Eprint {http://arxiv.org/abs/1303.1396} {arXiv:1303.1396 [hep-th]} \BibitemShut {NoStop}%
\bibitem [{\citenamefont {Basso}\ \emph {et~al.}(2014{\natexlab{a}})\citenamefont {Basso}, \citenamefont {Sever},\ and\ \citenamefont {Vieira}}]{Basso:2013aha}%
  \BibitemOpen
  \bibfield  {author} {\bibinfo {author} {\bibfnamefont {B.}~\bibnamefont {Basso}}, \bibinfo {author} {\bibfnamefont {A.}~\bibnamefont {Sever}}, \ and\ \bibinfo {author} {\bibfnamefont {P.}~\bibnamefont {Vieira}},\ }\href {\doibase 10.1007/JHEP01(2014)008} {\bibfield  {journal} {\bibinfo  {journal} {JHEP}\ }\textbf {\bibinfo {volume} {01}},\ \bibinfo {pages} {008} (\bibinfo {year} {2014}{\natexlab{a}})},\ \Eprint {http://arxiv.org/abs/1306.2058} {arXiv:1306.2058 [hep-th]} \BibitemShut {NoStop}%
\bibitem [{\citenamefont {Basso}\ \emph {et~al.}(2014{\natexlab{b}})\citenamefont {Basso}, \citenamefont {Sever},\ and\ \citenamefont {Vieira}}]{Basso:2014koa}%
  \BibitemOpen
  \bibfield  {author} {\bibinfo {author} {\bibfnamefont {B.}~\bibnamefont {Basso}}, \bibinfo {author} {\bibfnamefont {A.}~\bibnamefont {Sever}}, \ and\ \bibinfo {author} {\bibfnamefont {P.}~\bibnamefont {Vieira}},\ }\href {\doibase 10.1007/JHEP08(2014)085} {\bibfield  {journal} {\bibinfo  {journal} {JHEP}\ }\textbf {\bibinfo {volume} {08}},\ \bibinfo {pages} {085} (\bibinfo {year} {2014}{\natexlab{b}})},\ \Eprint {http://arxiv.org/abs/1402.3307} {arXiv:1402.3307 [hep-th]} \BibitemShut {NoStop}%
\bibitem [{\citenamefont {Basso}\ \emph {et~al.}(2015{\natexlab{a}})\citenamefont {Basso}, \citenamefont {Caetano}, \citenamefont {Cordova}, \citenamefont {Sever},\ and\ \citenamefont {Vieira}}]{Basso:2014hfa}%
  \BibitemOpen
  \bibfield  {author} {\bibinfo {author} {\bibfnamefont {B.}~\bibnamefont {Basso}}, \bibinfo {author} {\bibfnamefont {J.}~\bibnamefont {Caetano}}, \bibinfo {author} {\bibfnamefont {L.}~\bibnamefont {Cordova}}, \bibinfo {author} {\bibfnamefont {A.}~\bibnamefont {Sever}}, \ and\ \bibinfo {author} {\bibfnamefont {P.}~\bibnamefont {Vieira}},\ }\href {\doibase 10.1007/JHEP08(2015)018} {\bibfield  {journal} {\bibinfo  {journal} {JHEP}\ }\textbf {\bibinfo {volume} {08}},\ \bibinfo {pages} {018} (\bibinfo {year} {2015}{\natexlab{a}})},\ \Eprint {http://arxiv.org/abs/1412.1132} {arXiv:1412.1132 [hep-th]} \BibitemShut {NoStop}%
\bibitem [{\citenamefont {Basso}\ \emph {et~al.}(2015{\natexlab{b}})\citenamefont {Basso}, \citenamefont {Caetano}, \citenamefont {Cordova}, \citenamefont {Sever},\ and\ \citenamefont {Vieira}}]{Basso:2015rta}%
  \BibitemOpen
  \bibfield  {author} {\bibinfo {author} {\bibfnamefont {B.}~\bibnamefont {Basso}}, \bibinfo {author} {\bibfnamefont {J.}~\bibnamefont {Caetano}}, \bibinfo {author} {\bibfnamefont {L.}~\bibnamefont {Cordova}}, \bibinfo {author} {\bibfnamefont {A.}~\bibnamefont {Sever}}, \ and\ \bibinfo {author} {\bibfnamefont {P.}~\bibnamefont {Vieira}},\ }\href {\doibase 10.1007/JHEP12(2015)088} {\bibfield  {journal} {\bibinfo  {journal} {JHEP}\ }\textbf {\bibinfo {volume} {12}},\ \bibinfo {pages} {088} (\bibinfo {year} {2015}{\natexlab{b}})},\ \Eprint {http://arxiv.org/abs/1508.02987} {arXiv:1508.02987 [hep-th]} \BibitemShut {NoStop}%
\bibitem [{\citenamefont {Basso}\ \emph {et~al.}(2016)\citenamefont {Basso}, \citenamefont {Sever},\ and\ \citenamefont {Vieira}}]{Basso:2015uxa}%
  \BibitemOpen
  \bibfield  {author} {\bibinfo {author} {\bibfnamefont {B.}~\bibnamefont {Basso}}, \bibinfo {author} {\bibfnamefont {A.}~\bibnamefont {Sever}}, \ and\ \bibinfo {author} {\bibfnamefont {P.}~\bibnamefont {Vieira}},\ }\href {\doibase 10.1088/1751-8113/49/41/41LT01} {\bibfield  {journal} {\bibinfo  {journal} {J. Phys. A}\ }\textbf {\bibinfo {volume} {49}},\ \bibinfo {pages} {41LT01} (\bibinfo {year} {2016})},\ \Eprint {http://arxiv.org/abs/1508.03045} {arXiv:1508.03045 [hep-th]} \BibitemShut {NoStop}%
\bibitem [{\citenamefont {Belitsky}(2015{\natexlab{a}})}]{Belitsky:2014sla}%
  \BibitemOpen
  \bibfield  {author} {\bibinfo {author} {\bibfnamefont {A.~V.}\ \bibnamefont {Belitsky}},\ }\href {\doibase 10.1016/j.nuclphysb.2015.05.002} {\bibfield  {journal} {\bibinfo  {journal} {Nucl. Phys. B}\ }\textbf {\bibinfo {volume} {896}},\ \bibinfo {pages} {493} (\bibinfo {year} {2015}{\natexlab{a}})},\ \Eprint {http://arxiv.org/abs/1407.2853} {arXiv:1407.2853 [hep-th]} \BibitemShut {NoStop}%
\bibitem [{\citenamefont {Belitsky}(2015{\natexlab{b}})}]{Belitsky:2014lta}%
  \BibitemOpen
  \bibfield  {author} {\bibinfo {author} {\bibfnamefont {A.~V.}\ \bibnamefont {Belitsky}},\ }\href {\doibase 10.1016/j.nuclphysb.2015.02.025} {\bibfield  {journal} {\bibinfo  {journal} {Nucl. Phys. B}\ }\textbf {\bibinfo {volume} {894}},\ \bibinfo {pages} {108} (\bibinfo {year} {2015}{\natexlab{b}})},\ \Eprint {http://arxiv.org/abs/1410.2534} {arXiv:1410.2534 [hep-th]} \BibitemShut {NoStop}%
\bibitem [{\citenamefont {Belitsky}(2017)}]{Belitsky:2016vyq}%
  \BibitemOpen
  \bibfield  {author} {\bibinfo {author} {\bibfnamefont {A.~V.}\ \bibnamefont {Belitsky}},\ }\href {\doibase 10.1016/j.nuclphysb.2017.08.011} {\bibfield  {journal} {\bibinfo  {journal} {Nucl. Phys. B}\ }\textbf {\bibinfo {volume} {923}},\ \bibinfo {pages} {588} (\bibinfo {year} {2017})},\ \Eprint {http://arxiv.org/abs/1607.06555} {arXiv:1607.06555 [hep-th]} \BibitemShut {NoStop}%
\bibitem [{\citenamefont {Sever}\ \emph {et~al.}(2021{\natexlab{a}})\citenamefont {Sever}, \citenamefont {Tumanov},\ and\ \citenamefont {Wilhelm}}]{Sever:2020jjx}%
  \BibitemOpen
  \bibfield  {author} {\bibinfo {author} {\bibfnamefont {A.}~\bibnamefont {Sever}}, \bibinfo {author} {\bibfnamefont {A.~G.}\ \bibnamefont {Tumanov}}, \ and\ \bibinfo {author} {\bibfnamefont {M.}~\bibnamefont {Wilhelm}},\ }\href {\doibase 10.1103/PhysRevLett.126.031602} {\bibfield  {journal} {\bibinfo  {journal} {Phys. Rev. Lett.}\ }\textbf {\bibinfo {volume} {126}},\ \bibinfo {pages} {031602} (\bibinfo {year} {2021}{\natexlab{a}})},\ \Eprint {http://arxiv.org/abs/2009.11297} {arXiv:2009.11297 [hep-th]} \BibitemShut {NoStop}%
\bibitem [{\citenamefont {Sever}\ \emph {et~al.}(2021{\natexlab{b}})\citenamefont {Sever}, \citenamefont {Tumanov},\ and\ \citenamefont {Wilhelm}}]{Sever:2021nsq}%
  \BibitemOpen
  \bibfield  {author} {\bibinfo {author} {\bibfnamefont {A.}~\bibnamefont {Sever}}, \bibinfo {author} {\bibfnamefont {A.~G.}\ \bibnamefont {Tumanov}}, \ and\ \bibinfo {author} {\bibfnamefont {M.}~\bibnamefont {Wilhelm}},\ }\href {\doibase 10.1007/JHEP10(2021)071} {\bibfield  {journal} {\bibinfo  {journal} {JHEP}\ }\textbf {\bibinfo {volume} {10}},\ \bibinfo {pages} {071} (\bibinfo {year} {2021}{\natexlab{b}})},\ \Eprint {http://arxiv.org/abs/2105.13367} {arXiv:2105.13367 [hep-th]} \BibitemShut {NoStop}%
\bibitem [{\citenamefont {Sever}\ \emph {et~al.}(2022)\citenamefont {Sever}, \citenamefont {Tumanov},\ and\ \citenamefont {Wilhelm}}]{Sever:2021xga}%
  \BibitemOpen
  \bibfield  {author} {\bibinfo {author} {\bibfnamefont {A.}~\bibnamefont {Sever}}, \bibinfo {author} {\bibfnamefont {A.~G.}\ \bibnamefont {Tumanov}}, \ and\ \bibinfo {author} {\bibfnamefont {M.}~\bibnamefont {Wilhelm}},\ }\href {\doibase 10.1007/JHEP03(2022)128} {\bibfield  {journal} {\bibinfo  {journal} {JHEP}\ }\textbf {\bibinfo {volume} {03}},\ \bibinfo {pages} {128} (\bibinfo {year} {2022})},\ \Eprint {http://arxiv.org/abs/2112.10569} {arXiv:2112.10569 [hep-th]} \BibitemShut {NoStop}%
\bibitem [{\citenamefont {Basso}\ and\ \citenamefont {Tumanov}(2024)}]{Basso:2023bwv}%
  \BibitemOpen
  \bibfield  {author} {\bibinfo {author} {\bibfnamefont {B.}~\bibnamefont {Basso}}\ and\ \bibinfo {author} {\bibfnamefont {A.~G.}\ \bibnamefont {Tumanov}},\ }\href {\doibase 10.1007/JHEP02(2024)022} {\bibfield  {journal} {\bibinfo  {journal} {JHEP}\ }\textbf {\bibinfo {volume} {02}},\ \bibinfo {pages} {022} (\bibinfo {year} {2024})},\ \Eprint {http://arxiv.org/abs/2308.08432} {arXiv:2308.08432 [hep-th]} \BibitemShut {NoStop}%
\bibitem [{\citenamefont {Gaiotto}\ \emph {et~al.}(2011)\citenamefont {Gaiotto}, \citenamefont {Maldacena}, \citenamefont {Sever},\ and\ \citenamefont {Vieira}}]{Gaiotto:2011dt}%
  \BibitemOpen
  \bibfield  {author} {\bibinfo {author} {\bibfnamefont {D.}~\bibnamefont {Gaiotto}}, \bibinfo {author} {\bibfnamefont {J.}~\bibnamefont {Maldacena}}, \bibinfo {author} {\bibfnamefont {A.}~\bibnamefont {Sever}}, \ and\ \bibinfo {author} {\bibfnamefont {P.}~\bibnamefont {Vieira}},\ }\href {\doibase 10.1007/JHEP12(2011)011} {\bibfield  {journal} {\bibinfo  {journal} {JHEP}\ }\textbf {\bibinfo {volume} {12}},\ \bibinfo {pages} {011} (\bibinfo {year} {2011})},\ \Eprint {http://arxiv.org/abs/1102.0062} {arXiv:1102.0062 [hep-th]} \BibitemShut {NoStop}%
\bibitem [{\citenamefont {Dixon}\ \emph {et~al.}(2011)\citenamefont {Dixon}, \citenamefont {Drummond},\ and\ \citenamefont {Henn}}]{Dixon:2011pw}%
  \BibitemOpen
  \bibfield  {author} {\bibinfo {author} {\bibfnamefont {L.~J.}\ \bibnamefont {Dixon}}, \bibinfo {author} {\bibfnamefont {J.~M.}\ \bibnamefont {Drummond}}, \ and\ \bibinfo {author} {\bibfnamefont {J.~M.}\ \bibnamefont {Henn}},\ }\href {\doibase 10.1007/JHEP11(2011)023} {\bibfield  {journal} {\bibinfo  {journal} {JHEP}\ }\textbf {\bibinfo {volume} {11}},\ \bibinfo {pages} {023} (\bibinfo {year} {2011})},\ \Eprint {http://arxiv.org/abs/1108.4461} {arXiv:1108.4461 [hep-th]} \BibitemShut {NoStop}%
\bibitem [{\citenamefont {Dixon}\ \emph {et~al.}(2013)\citenamefont {Dixon}, \citenamefont {Drummond}, \citenamefont {von Hippel},\ and\ \citenamefont {Pennington}}]{Dixon:2013eka}%
  \BibitemOpen
  \bibfield  {author} {\bibinfo {author} {\bibfnamefont {L.~J.}\ \bibnamefont {Dixon}}, \bibinfo {author} {\bibfnamefont {J.~M.}\ \bibnamefont {Drummond}}, \bibinfo {author} {\bibfnamefont {M.}~\bibnamefont {von Hippel}}, \ and\ \bibinfo {author} {\bibfnamefont {J.}~\bibnamefont {Pennington}},\ }\href {\doibase 10.1007/JHEP12(2013)049} {\bibfield  {journal} {\bibinfo  {journal} {JHEP}\ }\textbf {\bibinfo {volume} {12}},\ \bibinfo {pages} {049} (\bibinfo {year} {2013})},\ \Eprint {http://arxiv.org/abs/1308.2276} {arXiv:1308.2276 [hep-th]} \BibitemShut {NoStop}%
\bibitem [{\citenamefont {Dixon}\ \emph {et~al.}(2014)\citenamefont {Dixon}, \citenamefont {Drummond}, \citenamefont {Duhr},\ and\ \citenamefont {Pennington}}]{Dixon:2014voa}%
  \BibitemOpen
  \bibfield  {author} {\bibinfo {author} {\bibfnamefont {L.~J.}\ \bibnamefont {Dixon}}, \bibinfo {author} {\bibfnamefont {J.~M.}\ \bibnamefont {Drummond}}, \bibinfo {author} {\bibfnamefont {C.}~\bibnamefont {Duhr}}, \ and\ \bibinfo {author} {\bibfnamefont {J.}~\bibnamefont {Pennington}},\ }\href {\doibase 10.1007/JHEP06(2014)116} {\bibfield  {journal} {\bibinfo  {journal} {JHEP}\ }\textbf {\bibinfo {volume} {06}},\ \bibinfo {pages} {116} (\bibinfo {year} {2014})},\ \Eprint {http://arxiv.org/abs/1402.3300} {arXiv:1402.3300 [hep-th]} \BibitemShut {NoStop}%
\bibitem [{\citenamefont {Dixon}\ and\ \citenamefont {von Hippel}(2014)}]{Dixon:2014iba}%
  \BibitemOpen
  \bibfield  {author} {\bibinfo {author} {\bibfnamefont {L.~J.}\ \bibnamefont {Dixon}}\ and\ \bibinfo {author} {\bibfnamefont {M.}~\bibnamefont {von Hippel}},\ }\href {\doibase 10.1007/JHEP10(2014)065} {\bibfield  {journal} {\bibinfo  {journal} {JHEP}\ }\textbf {\bibinfo {volume} {10}},\ \bibinfo {pages} {065} (\bibinfo {year} {2014})},\ \Eprint {http://arxiv.org/abs/1408.1505} {arXiv:1408.1505 [hep-th]} \BibitemShut {NoStop}%
\bibitem [{\citenamefont {Dixon}\ \emph {et~al.}(2016)\citenamefont {Dixon}, \citenamefont {von Hippel},\ and\ \citenamefont {McLeod}}]{Dixon:2015iva}%
  \BibitemOpen
  \bibfield  {author} {\bibinfo {author} {\bibfnamefont {L.~J.}\ \bibnamefont {Dixon}}, \bibinfo {author} {\bibfnamefont {M.}~\bibnamefont {von Hippel}}, \ and\ \bibinfo {author} {\bibfnamefont {A.~J.}\ \bibnamefont {McLeod}},\ }\href {\doibase 10.1007/JHEP01(2016)053} {\bibfield  {journal} {\bibinfo  {journal} {JHEP}\ }\textbf {\bibinfo {volume} {01}},\ \bibinfo {pages} {053} (\bibinfo {year} {2016})},\ \Eprint {http://arxiv.org/abs/1509.08127} {arXiv:1509.08127 [hep-th]} \BibitemShut {NoStop}%
\bibitem [{\citenamefont {Caron-Huot}\ \emph {et~al.}(2019)\citenamefont {Caron-Huot}, \citenamefont {Dixon}, \citenamefont {Dulat}, \citenamefont {von Hippel}, \citenamefont {McLeod},\ and\ \citenamefont {Papathanasiou}}]{Caron-Huot:2019vjl}%
  \BibitemOpen
  \bibfield  {author} {\bibinfo {author} {\bibfnamefont {S.}~\bibnamefont {Caron-Huot}}, \bibinfo {author} {\bibfnamefont {L.~J.}\ \bibnamefont {Dixon}}, \bibinfo {author} {\bibfnamefont {F.}~\bibnamefont {Dulat}}, \bibinfo {author} {\bibfnamefont {M.}~\bibnamefont {von Hippel}}, \bibinfo {author} {\bibfnamefont {A.~J.}\ \bibnamefont {McLeod}}, \ and\ \bibinfo {author} {\bibfnamefont {G.}~\bibnamefont {Papathanasiou}},\ }\href {\doibase 10.1007/JHEP08(2019)016} {\bibfield  {journal} {\bibinfo  {journal} {JHEP}\ }\textbf {\bibinfo {volume} {08}},\ \bibinfo {pages} {016} (\bibinfo {year} {2019})},\ \Eprint {http://arxiv.org/abs/1903.10890} {arXiv:1903.10890 [hep-th]} \BibitemShut {NoStop}%
\bibitem [{\citenamefont {Dixon}\ \emph {et~al.}(2021)\citenamefont {Dixon}, \citenamefont {McLeod},\ and\ \citenamefont {Wilhelm}}]{Dixon:2020bbt}%
  \BibitemOpen
  \bibfield  {author} {\bibinfo {author} {\bibfnamefont {L.~J.}\ \bibnamefont {Dixon}}, \bibinfo {author} {\bibfnamefont {A.~J.}\ \bibnamefont {McLeod}}, \ and\ \bibinfo {author} {\bibfnamefont {M.}~\bibnamefont {Wilhelm}},\ }\href {\doibase 10.1007/JHEP04(2021)147} {\bibfield  {journal} {\bibinfo  {journal} {JHEP}\ }\textbf {\bibinfo {volume} {04}},\ \bibinfo {pages} {147} (\bibinfo {year} {2021})},\ \Eprint {http://arxiv.org/abs/2012.12286} {arXiv:2012.12286 [hep-th]} \BibitemShut {NoStop}%
\bibitem [{\citenamefont {Dixon}\ \emph {et~al.}(2022)\citenamefont {Dixon}, \citenamefont {Gurdogan}, \citenamefont {McLeod},\ and\ \citenamefont {Wilhelm}}]{Dixon:2022rse}%
  \BibitemOpen
  \bibfield  {author} {\bibinfo {author} {\bibfnamefont {L.~J.}\ \bibnamefont {Dixon}}, \bibinfo {author} {\bibfnamefont {O.}~\bibnamefont {Gurdogan}}, \bibinfo {author} {\bibfnamefont {A.~J.}\ \bibnamefont {McLeod}}, \ and\ \bibinfo {author} {\bibfnamefont {M.}~\bibnamefont {Wilhelm}},\ }\href {\doibase 10.1007/JHEP07(2022)153} {\bibfield  {journal} {\bibinfo  {journal} {JHEP}\ }\textbf {\bibinfo {volume} {07}},\ \bibinfo {pages} {153} (\bibinfo {year} {2022})},\ \Eprint {http://arxiv.org/abs/2204.11901} {arXiv:2204.11901 [hep-th]} \BibitemShut {NoStop}%
\bibitem [{\citenamefont {Basso}\ \emph {et~al.}(2025)\citenamefont {Basso}, \citenamefont {Dixon},\ and\ \citenamefont {Tumanov}}]{Basso:2024hlx}%
  \BibitemOpen
  \bibfield  {author} {\bibinfo {author} {\bibfnamefont {B.}~\bibnamefont {Basso}}, \bibinfo {author} {\bibfnamefont {L.~J.}\ \bibnamefont {Dixon}}, \ and\ \bibinfo {author} {\bibfnamefont {A.~G.}\ \bibnamefont {Tumanov}},\ }\href {\doibase 10.1007/JHEP02(2025)034} {\bibfield  {journal} {\bibinfo  {journal} {JHEP}\ }\textbf {\bibinfo {volume} {02}},\ \bibinfo {pages} {034} (\bibinfo {year} {2025})},\ \Eprint {http://arxiv.org/abs/2410.22402} {arXiv:2410.22402 [hep-th]} \BibitemShut {NoStop}%
\bibitem [{\citenamefont {Remiddi}\ and\ \citenamefont {Vermaseren}(2000)}]{Remiddi:1999ew}%
  \BibitemOpen
  \bibfield  {author} {\bibinfo {author} {\bibfnamefont {E.}~\bibnamefont {Remiddi}}\ and\ \bibinfo {author} {\bibfnamefont {J.~A.~M.}\ \bibnamefont {Vermaseren}},\ }\href {\doibase 10.1142/S0217751X00000367} {\bibfield  {journal} {\bibinfo  {journal} {Int. J. Mod. Phys. A}\ }\textbf {\bibinfo {volume} {15}},\ \bibinfo {pages} {725} (\bibinfo {year} {2000})},\ \Eprint {http://arxiv.org/abs/hep-ph/9905237} {arXiv:hep-ph/9905237} \BibitemShut {NoStop}%
\bibitem [{\citenamefont {Alday}\ and\ \citenamefont {Maldacena}(2007{\natexlab{b}})}]{Alday:2007mf}%
  \BibitemOpen
  \bibfield  {author} {\bibinfo {author} {\bibfnamefont {L.~F.}\ \bibnamefont {Alday}}\ and\ \bibinfo {author} {\bibfnamefont {J.~M.}\ \bibnamefont {Maldacena}},\ }\href {\doibase 10.1088/1126-6708/2007/11/019} {\bibfield  {journal} {\bibinfo  {journal} {JHEP}\ }\textbf {\bibinfo {volume} {11}},\ \bibinfo {pages} {019} (\bibinfo {year} {2007}{\natexlab{b}})},\ \Eprint {http://arxiv.org/abs/0708.0672} {arXiv:0708.0672 [hep-th]} \BibitemShut {NoStop}%
\bibitem [{\citenamefont {Gubser}\ \emph {et~al.}(2002)\citenamefont {Gubser}, \citenamefont {Klebanov},\ and\ \citenamefont {Polyakov}}]{Gubser:2002tv}%
  \BibitemOpen
  \bibfield  {author} {\bibinfo {author} {\bibfnamefont {S.~S.}\ \bibnamefont {Gubser}}, \bibinfo {author} {\bibfnamefont {I.~R.}\ \bibnamefont {Klebanov}}, \ and\ \bibinfo {author} {\bibfnamefont {A.~M.}\ \bibnamefont {Polyakov}},\ }\href {\doibase 10.1016/S0550-3213(02)00373-5} {\bibfield  {journal} {\bibinfo  {journal} {Nucl. Phys. B}\ }\textbf {\bibinfo {volume} {636}},\ \bibinfo {pages} {99} (\bibinfo {year} {2002})},\ \Eprint {http://arxiv.org/abs/hep-th/0204051} {arXiv:hep-th/0204051} \BibitemShut {NoStop}%
\bibitem [{\citenamefont {Belitsky}\ \emph {et~al.}(2006)\citenamefont {Belitsky}, \citenamefont {Gorsky},\ and\ \citenamefont {Korchemsky}}]{Belitsky:2006en}%
  \BibitemOpen
  \bibfield  {author} {\bibinfo {author} {\bibfnamefont {A.~V.}\ \bibnamefont {Belitsky}}, \bibinfo {author} {\bibfnamefont {A.~S.}\ \bibnamefont {Gorsky}}, \ and\ \bibinfo {author} {\bibfnamefont {G.~P.}\ \bibnamefont {Korchemsky}},\ }\href {\doibase 10.1016/j.nuclphysb.2006.04.030} {\bibfield  {journal} {\bibinfo  {journal} {Nucl. Phys. B}\ }\textbf {\bibinfo {volume} {748}},\ \bibinfo {pages} {24} (\bibinfo {year} {2006})},\ \Eprint {http://arxiv.org/abs/hep-th/0601112} {arXiv:hep-th/0601112} \BibitemShut {NoStop}%
\bibitem [{\citenamefont {Gehrmann}\ \emph {et~al.}(2016)\citenamefont {Gehrmann}, \citenamefont {Henn},\ and\ \citenamefont {Lo~Presti}}]{Gehrmann:2015bfy}%
  \BibitemOpen
  \bibfield  {author} {\bibinfo {author} {\bibfnamefont {T.}~\bibnamefont {Gehrmann}}, \bibinfo {author} {\bibfnamefont {J.~M.}\ \bibnamefont {Henn}}, \ and\ \bibinfo {author} {\bibfnamefont {N.~A.}\ \bibnamefont {Lo~Presti}},\ }\href {\doibase 10.1103/PhysRevLett.116.062001} {\bibfield  {journal} {\bibinfo  {journal} {Phys. Rev. Lett.}\ }\textbf {\bibinfo {volume} {116}},\ \bibinfo {pages} {062001} (\bibinfo {year} {2016})},\ \bibinfo {note} {[Erratum: Phys.Rev.Lett. 116, 189903 (2016)]},\ \Eprint {http://arxiv.org/abs/1511.05409} {arXiv:1511.05409 [hep-ph]} \BibitemShut {NoStop}%
\bibitem [{\citenamefont {Gehrmann}\ \emph {et~al.}(2018)\citenamefont {Gehrmann}, \citenamefont {Henn},\ and\ \citenamefont {Lo~Presti}}]{Gehrmann:2018yef}%
  \BibitemOpen
  \bibfield  {author} {\bibinfo {author} {\bibfnamefont {T.}~\bibnamefont {Gehrmann}}, \bibinfo {author} {\bibfnamefont {J.~M.}\ \bibnamefont {Henn}}, \ and\ \bibinfo {author} {\bibfnamefont {N.~A.}\ \bibnamefont {Lo~Presti}},\ }\href {\doibase 10.1007/JHEP10(2018)103} {\bibfield  {journal} {\bibinfo  {journal} {JHEP}\ }\textbf {\bibinfo {volume} {10}},\ \bibinfo {pages} {103} (\bibinfo {year} {2018})},\ \Eprint {http://arxiv.org/abs/1807.09812} {arXiv:1807.09812 [hep-ph]} \BibitemShut {NoStop}%
\bibitem [{\citenamefont {Chicherin}\ and\ \citenamefont {Sotnikov}(2020)}]{Chicherin:2020oor}%
  \BibitemOpen
  \bibfield  {author} {\bibinfo {author} {\bibfnamefont {D.}~\bibnamefont {Chicherin}}\ and\ \bibinfo {author} {\bibfnamefont {V.}~\bibnamefont {Sotnikov}},\ }\href {\doibase 10.1007/JHEP12(2020)167} {\bibfield  {journal} {\bibinfo  {journal} {JHEP}\ }\textbf {\bibinfo {volume} {20}},\ \bibinfo {pages} {167} (\bibinfo {year} {2020})},\ \Eprint {http://arxiv.org/abs/2009.07803} {arXiv:2009.07803 [hep-ph]} \BibitemShut {NoStop}%
\bibitem [{\citenamefont {Wu}\ \emph {et~al.}(2024)\citenamefont {Wu}, \citenamefont {Boehm}, \citenamefont {Ma}, \citenamefont {Xu},\ and\ \citenamefont {Zhang}}]{Wu:2023upw}%
  \BibitemOpen
  \bibfield  {author} {\bibinfo {author} {\bibfnamefont {Z.}~\bibnamefont {Wu}}, \bibinfo {author} {\bibfnamefont {J.}~\bibnamefont {Boehm}}, \bibinfo {author} {\bibfnamefont {R.}~\bibnamefont {Ma}}, \bibinfo {author} {\bibfnamefont {H.}~\bibnamefont {Xu}}, \ and\ \bibinfo {author} {\bibfnamefont {Y.}~\bibnamefont {Zhang}},\ }\href {\doibase 10.1016/j.cpc.2023.108999} {\bibfield  {journal} {\bibinfo  {journal} {Comput. Phys. Commun.}\ }\textbf {\bibinfo {volume} {295}},\ \bibinfo {pages} {108999} (\bibinfo {year} {2024})},\ \Eprint {http://arxiv.org/abs/2305.08783} {arXiv:2305.08783 [hep-ph]} \BibitemShut {NoStop}%
\bibitem [{\citenamefont {Wu}\ and\ \citenamefont {Zhang}(2024)}]{Wu:2024paw}%
  \BibitemOpen
  \bibfield  {author} {\bibinfo {author} {\bibfnamefont {Z.}~\bibnamefont {Wu}}\ and\ \bibinfo {author} {\bibfnamefont {Y.}~\bibnamefont {Zhang}},\ }\href@noop {} {\  (\bibinfo {year} {2024})},\ \Eprint {http://arxiv.org/abs/2406.20016} {arXiv:2406.20016 [hep-ph]} \BibitemShut {NoStop}%
\bibitem [{\citenamefont {Wu}\ \emph {et~al.}(2025)\citenamefont {Wu}, \citenamefont {B{\"o}hm}, \citenamefont {Ma}, \citenamefont {Usovitsch}, \citenamefont {Xu},\ and\ \citenamefont {Zhang}}]{Wu:2025aeg}%
  \BibitemOpen
  \bibfield  {author} {\bibinfo {author} {\bibfnamefont {Z.}~\bibnamefont {Wu}}, \bibinfo {author} {\bibfnamefont {J.}~\bibnamefont {B{\"o}hm}}, \bibinfo {author} {\bibfnamefont {R.}~\bibnamefont {Ma}}, \bibinfo {author} {\bibfnamefont {J.}~\bibnamefont {Usovitsch}}, \bibinfo {author} {\bibfnamefont {Y.}~\bibnamefont {Xu}}, \ and\ \bibinfo {author} {\bibfnamefont {Y.}~\bibnamefont {Zhang}},\ }\href {\doibase 10.1016/j.cpc.2025.109798} {\bibfield  {journal} {\bibinfo  {journal} {Comput. Phys. Commun.}\ }\textbf {\bibinfo {volume} {316}},\ \bibinfo {pages} {109798} (\bibinfo {year} {2025})},\ \Eprint {http://arxiv.org/abs/2502.20778} {arXiv:2502.20778 [hep-ph]} \BibitemShut {NoStop}%
\bibitem [{\citenamefont {Henn}\ and\ \citenamefont {Raman}(2025)}]{Henn:2024qwe}%
  \BibitemOpen
  \bibfield  {author} {\bibinfo {author} {\bibfnamefont {J.}~\bibnamefont {Henn}}\ and\ \bibinfo {author} {\bibfnamefont {P.}~\bibnamefont {Raman}},\ }\href {\doibase 10.1007/JHEP04(2025)150} {\bibfield  {journal} {\bibinfo  {journal} {JHEP}\ }\textbf {\bibinfo {volume} {04}},\ \bibinfo {pages} {150} (\bibinfo {year} {2025})},\ \Eprint {http://arxiv.org/abs/2407.05755} {arXiv:2407.05755 [hep-th]} \BibitemShut {NoStop}%
\bibitem [{\citenamefont {Chicherin}\ \emph {et~al.}(2018)\citenamefont {Chicherin}, \citenamefont {Henn},\ and\ \citenamefont {Mitev}}]{Chicherin:2017dob}%
  \BibitemOpen
  \bibfield  {author} {\bibinfo {author} {\bibfnamefont {D.}~\bibnamefont {Chicherin}}, \bibinfo {author} {\bibfnamefont {J.}~\bibnamefont {Henn}}, \ and\ \bibinfo {author} {\bibfnamefont {V.}~\bibnamefont {Mitev}},\ }\href {\doibase 10.1007/JHEP05(2018)164} {\bibfield  {journal} {\bibinfo  {journal} {JHEP}\ }\textbf {\bibinfo {volume} {05}},\ \bibinfo {pages} {164} (\bibinfo {year} {2018})},\ \Eprint {http://arxiv.org/abs/1712.09610} {arXiv:1712.09610 [hep-th]} \BibitemShut {NoStop}%
\end{thebibliography}%

\appendix
\widetext
\begin{center}
    \textbf{\large Supplemental Material}
\end{center}

\section{Alphabet letters appearing in the three-loop answer}
\label{app:letters}

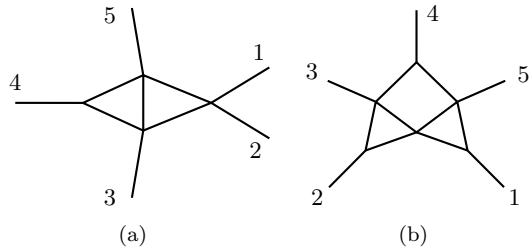
\begin{figure}[h]
\centering
			\subfigure[] {
				\begin{tikzpicture}[thick, scale=0.9]
			\coordinate (1) at (1.72324,0);
			\coordinate (2) at (0,1.40342);
			\coordinate (3) at (1.72325,2.80579);
			\coordinate (4) at (3.75371,0.878488);
			\coordinate (5) at (3.75369,1.92743);
			\coordinate (6) at (1.0008,1.40272);
			\coordinate (7) at (1.89078,0.993457);
			\coordinate (8) at (1.89099,1.81371);
			\coordinate (9) at (2.89417,1.40389);
			\draw (7)--(9)(6)--(8)(7)--(8)(6)--(7)(8)--(9);
			\draw (2)--(6) (1)--(7) (5)--(9) (3)--(8) (4)--(9);
			\node at  (1.4,0) {$3$};
			\node at  (0,1.7)    {$4$};
			\node at  (3.6,2.15) {$1$};
			\node at  (1.4, 2.7)  {$5$};
			\node at  (3.55,0.7)  {$2$};
				\end{tikzpicture}\label{fig:w21}
                }
\subfigure[] {
				\begin{tikzpicture}[thick, scale=0.7]
					\coordinate (1) at (0.0251276,0);
					\coordinate (2) at (0,2.02397);
					\coordinate (3) at (1.68607,3.36531);
					\coordinate (4) at (3.37098,2.0242);
					\coordinate (5) at (3.3484,0.00202879);
					\coordinate (6) at (0.71801,0.698364);
					\coordinate (7) at (1.68574,2.37617);
					\coordinate (8) at (2.65542,0.698376);
					\coordinate (9) at (0.912256,1.62474);
					\coordinate (10) at (2.46072,1.62462);
					\coordinate (11) at (1.68593,1.04422);
					\draw (6)--(9)(10)--(11)(7)--(9)(6)--(11)(7)--(10)(9)--(11)(8)--(10)(8)--(11);
					\draw (1)--(6) (3)--(7) (2)--(9) (4)--(10) (5)--(8);
					\node at  (-0.2,-0.2) {$2$};
					\node at  (-0.3,2.2)  {$3$};
					\node at  (3.72,2.15)  {$5$};
					\node at  (2,3.3)  {$4$};
					\node at  (3.55,-0.2)  {$1$};
				\end{tikzpicture}\label{fig:w26}
                }
\caption{Feynman integrals that give rise to the letters $w_{21}$ and $w_{26}$, respectively.
}
\label{fig:alphabetletters}
\end{figure}

For completeness, we list the $30$ symbol alphabet letters relevant for the three-loop observable. We denote them by $w_{i}$ with $1 \le i \le 30$.
They can be organized into cyclic orbits of five elements each. For brevity, below we give one of the elements of the orbits.
$20$ of the letters appear already at one loop, namely 
\begin{equation}
\begin{aligned}
w_1&=s_{12}\,, \\
w_{6}&=s_{12}-s_{45}\,, \\
w_{11}&=s_{12}+s_{23}-s_{45} \,,\\
\end{aligned}
\end{equation}
and 
\begin{align}
w_{16}=\frac{-s_{12} s_{15} + s_{12} s_{23} - s_{23} s_{34} - s_{15} s_{45} + s_{34} s_{45}-\epsilon_5}{-s_{12} s_{15} + s_{12} s_{23} - s_{23} s_{34} - s_{15} s_{45} + s_{34} s_{45}+\epsilon_5} \,,
\end{align}
where $\epsilon_5\equiv 4i\epsilon_{\mu_1 \mu_2 \mu_3 \mu_4} p_1^{\mu_1} p_2^{\mu_2} p_3^{\mu_3} p_4^{\mu_4}$.
They also have simple expressions in terms of spinor variables. In particular, we have
\begin{align}
    w_{16} = \frac{\langle45\rangle[51]\langle12\rangle[24]}{[45]\langle51\rangle[12]\langle24\rangle}  \,.
\end{align}
At two loops, $5$ more letters appear, namely
\begin{align}
w_{21}=s_{34}+s_{45} \,.
\end{align}
They can be associated to the Feynman diagram shown in Fig.~\ref{fig:w21}.
Here we have done a reordering w.r.t. the order of letters in ref. \cite{Chicherin:2017dob}.
They are organized into five cyclic orbits that are generated by $\tau$.

Our result shows that at three loops, only five additional letters are required. The latter were predicted in ref. \cite{Chicherin:2024hes}.
In particular, they can be seen to arise from 
the Feynman integral displayed in Fig.~\ref{fig:w26}.
They are given by
\begin{align}
w_{26} = s_{23} s_{34} - s_{34} s_{45} + s_{45} s_{15} \,,\label{eq:What1}
\end{align}
plus four letters obtained by cyclic permutations.

With this notation, the alphabet letters arising at different entries in the symbol of $F^{(3)}$, discussed in section \ref{sec:discussion}, are given in 
Table~\ref{tab:letters}. 

\begin{table}[t]
\begin{tabular}{ccc}
\toprule
Symbol entry  & Letters  & $\#$ of letters \\
\midrule
$1$ & $w_{1,...,5}$ & 5\\
$2$ & $w_{1,...,10}$ & 10 \\
$3$ & $w_{1,...,20}$ & 20\\
$4$ & $w_{1,...,30}$ &30 \\
$5$ & $w_{1,...,25}$ &25\\
$6$  & $w_{1,...,25}$  &25\\
\bottomrule
\end{tabular}
\caption{\label{tab:letters} Letters appearing in $F^{(3)}$ at different symbol entries.}
\end{table}

\end{document}